\documentclass[lettersize,11pt]{article}
\pdfoutput=1 

\usepackage{jheppub, colonequals} 

\usepackage[T1]{fontenc} 
\usepackage{subcaption}
\preprint{YITP-SB-2017-45}

\newcommand{\OO}{{\cal O}}

\title{\boldmath \Large
Holographic Four-Point Functions in the $(2, 0)$ Theory }

\author[]{Leonardo Rastelli,}
\author[]{Xinan Zhou}

\affiliation[]{C. N. Yang Institute for Theoretical Physics, Stony Brook University, Stony Brook, 11794, NY, USA}

\emailAdd{leonardo.rastelli@stonybrook.edu}
\emailAdd{xinan.zhou@stonybrook.edu}

\graphicspath{ {graphs/} }

\abstract{We revisit the calculation of holographic correlators for eleven-dimensional supergravity on  $AdS_7\times S^4$. Our methods rely entirely on symmetry and eschew detailed knowledge of the supergravity effective action.
By an extension of the  position space approach
developed in \cite{Rastelli:2016nze,longads5} for the $AdS_5\times S^5$ background, we compute four-point correlators of one-half BPS operators for identical  weights $k=2, 3, 4$. The $k=2$ case corresponds to the four-point function of the stress-tensor multiplet, which was already known, while the other two cases are new. 
We also translate the problem in Mellin space, where the solution of the superconformal Ward identity takes a surprisingly simple form. We formulate an algebraic problem, whose (conjecturally unique) solution corresponds to the general one-half BPS four-point function.
 }

\begin{document} 
\maketitle
\flushbottom
\section{Introduction }

Correlation functions of local operators in holographic CFTs were among the first observables to be computed using the AdS/CFT dictionary,
but only recently  \cite{Rastelli:2016nze,longads5} efficient calculational methods  have begun to be developed.
The traditional recipe 
is based on a perturbative expansion in Witten diagrams, which becomes very cumbersome (already at tree level) for $n$-point correlators with $n \geqslant 4$.
Prior to our work,  only  a few   four-point correlators of Kaluza-Klein (KK) modes
were known, to wit (focussing for definiteness on the maximally supersymmetric backgrounds): a handful of cases 
in the $AdS_5 \times S^5$ background \cite{Freedman:1998bj, Arutyunov:2000py, Arutyunov:2002fh, Arutyunov:2003ae, Berdichevsky:2007xd,Uruchurtu:2008kp,Uruchurtu:2011wh}; just  the four-point function of
the {\it lowest} KK mode (the stress-tensor multiplet) in the $AdS_7 \times S^4$ background \cite{Arutyunov:2002ff}; and no results whatsoever in the $AdS_4 \times S^7$ background.

\smallskip

The traditional method has two sources of computational complexity: the need for explicit expressions of the vertices in the supergravity effective action;
and the proliferation of exchange Witten diagrams as the KK level is increased.  In \cite{Rastelli:2016nze,longads5} we introduced new calculational tools  to  circumvent these difficulties, for the case of  $AdS_5 \times S^5$ supergravity.
A  first approach, which we refer to as the ``position space method'',  leverages the special feature of  the $AdS_5 \times S^5$ background  that all exchange Witten diagrams can be written as finite sums of contact diagrams.
One can then write an ansatz for the four-point correlator as a sum of contact diagrams, and determine their relative coefficients by imposing superconformal symmetry, with no need for a detailed knowledge of the effective supergravity action. 
While simpler than the standard perturbative recipe,
the position space method also quickly runs out of steam as the KK level is increased. What's worse, the answer takes a completely unintuitive form, with no simpe general pattern.
The second, more powerful approach of \cite{Rastelli:2016nze,longads5} uses the Mellin representation of conformal correlators \cite{Mack:2009mi,Penedones:2010ue}. Tree-level holographic correlators in $AdS_5 \times S^5$ are rational functions
of Mandelstam-like invariants,  with poles and residues controlled by OPE factorization, in close analogy with tree-level flat space scattering amplitudes. 
Superconformal symmetry is made manifest  by solving the superconformal Ward identity in terms of an ``auxiliary'' Mellin amplitude. The consistency conditions that this amplitude must satisfy define a very constrained algebraic problem,
which very plausibly admits a unique solution.    While the position space method is implemented  on a case-by-case basis for different correlators, the Mellin algebraic problem  takes a universal form. We were able to solve the problem in one fell swoop 
for all half-BPS four-point function with arbitrary  weights -- a feat extremely difficult to replicate in position space.

\smallskip

The goal of this note is to extend these techniques to  eleven-dimensional supergravity on $AdS_7\times S^4$.  This is a background of extraordinary physical interest, 
as it provides  a dual description of the  mysterious six-dimensional $(2,0)$ theory  at large $n$, and 
 a prime target for our methods, since we expect maximal supersymmetry to constrain the tree-level holographic correlators uniquely.
At a more technical level, this background enjoys the same ``truncation conditions'' as $AdS_5 \times S^5$, such that exchange Witten diagrams can be written as finite sums of contact diagrams -- equivalently, tree-level Mellin amplitudes
are {\it rational} functions of the Mandelstam invariants.  By contrast, the truncation conditions do not hold for the $AdS_4 \times S^7$ background, and new tools are needed \cite{Zhou}.

\smallskip
In this paper we focus on four-point functions of identical one-half BPS local operators,
\begin{equation}
\langle \mathcal{O}^{(k)}(x_1)\mathcal{O}^{(k)}(x_2)\mathcal{O}^{(k)}(x_3)\mathcal{O}^{(k)}(x_4)\rangle\;.
\end{equation}
The operator $\mathcal{O}^{(k)}$  
is the superconformal primary of the short multiplet\footnote{For the representation theory of $d=6$ $(2,0)$ superconformal algebra, see, {\it e.g.}, \cite{Minwalla:1997ka,Dobrev:2002dt} and the summary in \cite{Beem:2014kka} whose notations we follow. A general multiplet is denoted as $\mathcal{X}\big(\Delta;J_1,J_2,J_3;a_1,a_2\big)$, where $\Delta$ is the conformal dimension, $\{J_i\}$ label the irreducible representation of the Lorentz group and $[a_1,a_2]$ is the Dynkin label of the $USp(4)$ R-symmetry.} $\mathcal{D}\big(2k;0,0,0;0,k\big)$. It has conformal dimension $\Delta=2k$ and transforms in the symmetric traceless representation of the R-symmetry group $SO(5) \cong USp(4)$. The integer $k \geqslant 2$ corresponds to the Kaluza-Klein level on the supergravity side. In particular  $\mathcal{O}^{(2)}$ is the superprimary
of the stress tensor multiplet, whose four-point function has been computed by Arutyunov and Sokatchev  \cite{Arutyunov:2002ff}, while operators with $k\geqslant 3$ are dual to massive Kaluza-Klein modes. No four-point functions for massive KK modes
have ever been computed, partly because of the extraordinary challenge of evaluating the quartic vertices of the $AdS_7$ effective action from the non-linear KK reduction. 

\smallskip

The position space method of \cite{Rastelli:2016nze,longads5} admits a straightforward extension to the $AdS_7 \times S^4$ background.  Using this method we confirm the result of \cite{Arutyunov:2002ff} for the stress-tensor multiplet four-point function
and obtain two new explicit cases with KK levels $k=3$ and $k=4$.
 An important test of our results is that they are compatible with the expected chiral algebra structure identified in   \cite{Beem:2013sza,Beem:2014kka}.
By performing a certain ``twist'' that identifies spacetime and R-symmetry cross ratios, correlators of a $6d$ (2, 0) superconformal theory
map to correlators of a $2d$ chiral algebra. In \cite{Beem:2014kka}, the chiral algebra associated to the $(2, 0)$ theory of type $A_n$ was identified with the $W_n$ algebra with central charge $c_{2d} =  4 n^3-3n-1$. In particular, three-point functions
of one-half BPS operators, previously computed at large $n$ from supergravity, were shown to agree with the structure constants of $W_{n \to \infty}$. Here we check that our newly obtained holographic four-point functions reduce  to  four-point functions
of the $W_{n \to \infty}$ algebra upon performing the chiral algebra twist.\footnote{To perform this test, we need to independently compute the requisite meromorphic four-point functions of $W_{n \to \infty}$ by purely two-dimensional method. This is easily done by fixing  their poles and residues  from OPE factorization and requiring crossing symmetry. This calculation uses as an input the structure constants of $W_{n \to \infty}$, so compatibility of our  holographic four-point function with the $W_{n \to \infty}$
structure is not really an independent check of the identification  \cite{Beem:2014kka} of $W_n$ as the correct chiral algebra. It is however a very  non-trivial check of our computations.}

\smallskip

We also reformulate the problem in Mellin space. The same strategy of  \cite{Rastelli:2016nze,longads5} applies, but the $AdS_7$ case is significantly more involved. The solution 
of the superconformal Ward identity   takes a much more intricate form in position space, and its translation to Mellin space requires some non-trivial manipulations. Nevertheless, when the dust settles, we  find a surprisingly compact
way to express the constraints of superconformal invariance in Mellin space, structurally similar to the $AdS_5$ case. The Mellin amplitude can be written in terms of an auxiliary amplitude, 
 acted upon  by a certain difference operator. We  formulate a purely algebraic problem based on symmetries and consistency conditions that we believe should determine uniquely the Mellin amplitude, but
 unlike the $AdS_5$ case, we have not yet been able to conjecture a general solution. 
 We have translated our position space results 
 into Mellin space and found that the auxiliary Mellin amplitude takes a succinct form for $k=2$ and $k=3$, and already a rather complicated form for $k=4$ (even if significantly more compact than the position space expression).
 We were not able to guess a pattern for the amplitude for general $k$. This remains the main outstanding challenge. If that can be overcome, we can look forward to an extension to the $AdS_7 \times S^4$ background of the  ideas and techniques
 that have been recently applied to the $AdS_5 \times S^5$ case, see, {\it e.g.},  \cite{Alday:2017gde,Alday:2017xua,Aprile:2017xsp,Aprile:2017bgs,Alday:2017vkk, Aprile:2017qoy}.

The remainder of the paper is  organized as follows.  In Section \ref{skine}  we set up  notations and review the superconformal kinematics of  one-half BPS four-point functions. We formulate the position space method in Section \ref{pmethod} and present the results of the first three low-lying correlators in the subsequent subsections. In Section \ref{manifestscf} we translate the constraints of superconformal symmetry into Mellin space. Using this formalism we formulate  in Section \ref{algeprob} an algebraic problem that is expected
to completely determine the  one-half BPS four-point functions.
 In Section \ref{mamplitudes} we present the Mellin space translation of our
 three position space solutions. Technical details (including a computation of $\mathcal{W}_{n \to \infty}$ four-point functions) are relegated to the three appendices.

\section{Superconformal kinematics of four-point functions}\label{skine}

We begin by reviewing the kinematic
constraints of superconformal invariance on one-half BPS four-point functions.  The relevant superalgebra 
$\mathfrak{osp}(8^*|4)$, whose bosonic subalgebra is the direct sum of the conformal algebra $\mathfrak{so}(6, 2)$ and  of the 
 R-symmetry algebra $\mathfrak{so}(5)$.

  One-half BPS operators ${\cal O}^{(k)}_{I_1\ldots I_k}$ transform in the symmetric traceless representation of the $\mathfrak{so}(5)$ $R$-symmetry. They can be conveniently made index-free by contracting the $\mathfrak{so}(5)$  indices with five-dimensional null vectors $t^I$,
\begin{equation}
\OO^{(k)}(x,t)=\OO^{(k)}_{I_1\ldots I_k}(x)t^{I_1}\ldots t^{I_k}\;,\;\;\;\;\;\;\;\; t^2=0\;.
\end{equation}
The four-point function
\begin{equation}
G_k(x_i,t_i)=\langle \OO^{(k)}(x_1,t_1)\OO^{(k)}(x_2,t_2)\OO^{(k)}(x_3,t_3)\OO^{(k)}(x_4,t_4)\rangle
\end{equation}
is thus  a function of the spacetime coordinates $x_i$ and of the ``internal'' coordinates $t_i$. Covariance under  $\mathfrak{so}(5)$ and $\mathfrak{so}(6,2)$
can be exploited to pull out a kinematic factor,
\begin{equation}\label{GUVst}
G_k(x_i,t_i)=\left(\frac{t_{12}\; t_{34}}{x_{12}^4x_{34}^4}\right)^k\mathcal{G}_k(U,V;\sigma,\tau) \, ,
\end{equation}
such that the reduced correlator $\mathcal{G}_k$ depends only on the conformal  cross ratios $U$ and $V$ and on the R-symmetry cross ratios $\sigma$ and $\tau$. The cross ratios have the standard definitions
\begin{equation}
      U =  \frac{(x_{12})^2(x_{34})^2}{(x_{13})^2(x_{24})^2}\;,\quad\quad V =   \frac{(x_{14})^2(x_{23})^2}{(x_{13})^2(x_{24})^2}\;,\quad\quad x_{ij}\colonequals x_i-x_j
\end{equation} 
and 
\begin{equation}
       \sigma = \frac{(t_{13}) (t_{24})}{(t_{12}) (t_{34})}\;,\quad\quad
       \tau=  \frac{(t_{14}) (t_{23})}{(t_{12}) (t_{34})}\;,\quad\quad t_{ij}\colon= t_i\cdot t_j\;.
\end{equation}
The  correlator  $G_k(x_i,t_i)$ is a polynomial 
in the R-symmetry invariants $t_{ij}$. 
 It is  easy to see that the reduced correlator $\mathcal{G}_k(U,V;\sigma,\tau)$ is a degree-$k$ polynomial of $\sigma$ and $\tau$.

So far we have only imposed the constraints that arise from the bosonic subalgebra of the full superalgebra. The fermionic generators imply  further constraints, which 
 link the dependence  on the R-symmetry and conformal cross ratios. After making the convenient change of variables 
\begin{equation}\label{changevariable}
\begin{split}
{}&U=\chi\chi'\;, \;\;\;\;\;\;V=(1-\chi)(1-\chi')\;,\\
{}&\sigma=\alpha\alpha'\;, \;\;\;\;\;\;\tau=(1-\alpha)(1-\alpha')\;,
\end{split}
\end{equation}
the superconformal Ward identity reads  \cite{Dolan:2004mu}
\begin{equation}\label{scfwi}
(\chi'\partial_{\chi'}-2\alpha'\partial_{\alpha'})\mathcal{G}_k(\chi,\chi';\alpha,\alpha')\big|_{\alpha'\to1/\chi'}=0\;.
\end{equation}
We can first obtain a partial solution to the superconformal Ward identity by restricting the four-point function to a special slice of R-symmetry cross ratios such that $\alpha=\alpha'=1/\chi'$. Then the superconformal Ward identity (\ref{scfwi}) reduces to
\begin{equation}
\chi'\partial_{\chi'}\mathcal{G}_k(\chi,\chi';1/\chi',1/\chi')=0\; ,
\end{equation}
whose solution is simply any ``holomorphic'' function\footnote{In Euclidean signature, the variables $\chi$ and $\chi'$ are complex conjugate of each other, so this terminology is appropriate. 
In Lorentzian signature $\chi$ and $\chi'$ are instead real independent variables. Hence ``holomorphic'' in quotes.} of $\chi$,
 \begin{equation}
 \mathcal{G}_k(\chi,\chi';1/\chi',1/\chi')=f(\chi)\;.
 \end{equation}
Up to kinematic factors, the  function  $f(\chi)$ coincides with the four-point correlator of the two-dimensional chiral algebra  associated to the $(2, 0)$ theory by the cohomological procedure  introduced in 
\cite{Beem:2013sza,Beem:2014kka}. There is a compelling conjecture \cite{Beem:2014kka} that  the chiral algebra associated to the $(2, 0)$ theory of type $A_n$ is the  familiar $\mathcal{W}_n$ algebra, with central charge $c_{2d} =4 n^3-3n-1$.
In our holographic setting, we are instructed to take a suitable large $n$ limit of the  $\mathcal{W}_n$ algebra, as explained in detail in  \cite{Beem:2014kka}.  In that limit, the structure constants of the $\mathcal{W}_n$
algebra were matched with the three-point functions of the one-half BPS operators computed holographically by standard supergravity methods.
In this work, we will use our new method to compute holographic {\it four}-point functions of one-half BPS operators.
As an important consistency check, we will  match their  ``holomorphic'' piece $f(\chi)$  with the corresponding four-point functions in the $\mathcal{W}_{n \to \infty}$ algebra.

The full solution of the superconformal Ward identity (\ref{scfwi}) was found in  \cite{Dolan:2004mu}. We reproduce it here with a few crucial typos fixed. A general solution $\mathcal{G}_k$ of  (\ref{scfwi})  can be written as
\begin{equation}\label{scfwisol}
\mathcal{G}_k(U,V;\sigma,\tau)=\mathcal{F}_k(U,V;\sigma,\tau)+\mathcal{K}_k(U,V;\sigma,\tau)\; ,
\end{equation}
where  $\mathcal{F}_k$ and $\mathcal{K}_k$ are respectively an ``inhomogeneous'' solution'' and a ``homogenous'' solution.  By this we mean that upon performing the  ``twist'' $\alpha=\alpha'=1/\chi'$, $\mathcal{F}_k$ becomes a purely ``holomorphic'' function 
of $\chi$, while $\mathcal{K}_k$ must vanish identically. The homogenous part $\mathcal{K}_k$ can further be expressed in terms of a differential operator $\Upsilon$ acting on an unconstrained function $\mathcal{H}(U,V;\sigma,\tau)$,
 which is a polynomial in $\sigma$ and $\tau$ of degree $k-2$. Explicitly,\footnote{Here and below, the parameter $\epsilon$ takes the fixed value 2. We keep it as $\epsilon$  to facilitate comparing with the expressions in \cite{Dolan:2004mu}, but we stress that the solution to the superconformal Ward identity  takes this particular form only for $d=6$.}
\begin{equation}\label{upsilon}
\begin{split}
\mathcal{K}_k(U,V;\sigma,\tau)={}&\big(\sigma^2\mathcal{D}'_{\epsilon} UV+\tau^2 \mathcal{D}'_{\epsilon} U+\mathcal{D}'_{\epsilon}V-\sigma \mathcal{D}'_{\epsilon} V(U+1-V)\\
-{}&\tau \mathcal{D}'_{\epsilon} (U+V-1)-\sigma\tau \mathcal{D}'_{\epsilon} U(V+1-U)\big)\mathcal{H}_k(U,V;\sigma,\tau)\\
\equalscolon& \Upsilon \circ \mathcal{H}_k(U,V;\sigma,\tau)\; ,
\end{split}
\end{equation}
where the
differential operator $\mathcal{D}'_\epsilon$ is defined as
\begin{eqnarray}
\mathcal{D}'_\epsilon &\colonequals & \bigg[D_\epsilon-\frac{\epsilon}{V}(D_0^+-D_1^++\epsilon\partial_\sigma\sigma)\tau\partial_\tau+\frac{\epsilon}{UV}(-VD_1^++\epsilon(V\partial_\sigma\sigma+\partial_\tau\tau-1))(\partial_\sigma\sigma+\partial_\tau\tau)\bigg]^{\epsilon-1}\, ,\nonumber \\
D_\epsilon & \colonequals &\frac{\partial^2}{\partial\chi\partial\chi'}-\epsilon\frac{1}{\chi-\chi'}(\frac{\partial}{\partial\chi}-\frac{\partial}{\partial\chi'})\;,\\
D_0^+ & \colonequals &\frac{\partial}{\partial \chi}+\frac{\partial}{\partial \chi'}\;,  \nonumber \\
D_1^+ & \colonequals &\chi\frac{\partial}{\partial \chi}+\chi'\frac{\partial}{\partial \chi'}\;. \nonumber
\end{eqnarray}
While the expression of the differential operator $\Upsilon$ is not very transparent, its transformation properties under crossing however are surprisingly simple. Let $g_1$, $g_2$ be the two generators of the crossing-symmetry group $S_3$ under which the cross ratios transform as
\begin{equation}
\begin{split}
g_1:{}&\;\;\;\;\;\;\; U\to \frac{U}{V}\;,\;\;V\to \frac{1}{V} \;,\;\; \sigma\to\tau\;,\;\;\tau\to\sigma\;,\\
g_2:{}&\;\;\;\;\;\;\; U\to\frac{1}{U}\;,\;\;V\to \frac{V}{U} \;,\;\; \sigma\to \frac{1}{\sigma}\;,\;\;\tau \to \frac{\tau}{\sigma}\;.
\end{split}
\end{equation}
We have found that  $\Upsilon$ satisfies\footnote{It is understood here that both sides of  (\ref{upsiloncross}) are acting on the same arbitrary function of the cross-ratios.}
\begin{equation}\label{upsiloncross}
\begin{split}
g_1\circ\Upsilon={}& \sigma ^{\frac{1}{2} (-\gamma -\rho )}  \tau ^{\rho /2} U^{-\gamma } V^{-\rho }\Upsilon\sigma ^{\frac{\gamma +\rho }{2}}\tau ^{-\frac{\rho }{2}} U^{\gamma } V^{\rho }\;,\\
g_2\circ\Upsilon={}& \sigma ^{\rho-\frac{\gamma }{2}} \tau ^{-\rho} V^{2 \rho} U^{-\gamma } \Upsilon\sigma ^{\frac{\gamma }{2}-\rho-2}\tau ^\rho V^{-2 \rho} U^{\gamma } \, ,
\end{split}
\end{equation}
where $\gamma$ and $\rho$ are arbitrary parameters.

We can always find a decomposition of $\mathcal{G}_k$  such that the two functions $\mathcal{F}_k$ and $\mathcal{H}_k$ do not mix into each other under crossing. Then the full correlator $\mathcal{G}_k$ and and the inhomogenous part  $\mathcal{F}_k$ have the same crossing properties
\begin{equation}
\begin{split}\label{crossingH}
\mathcal{G}_k(U,V;\sigma,\tau)={}&\left(\frac{U^2\tau}{V^2}\right)^k\mathcal{G}_{k}(V,U;\sigma/\tau,1/\tau)=\left(U^2\sigma \right)^k\mathcal{G}_{k}(1/U,V/U;1/\sigma,\tau/\sigma)\;,\\
\mathcal{F}_k(U,V;\sigma,\tau)={}&\left(\frac{U^2\tau}{V^2}\right)^k\mathcal{F}_{k}(V,U;\sigma/\tau,1/\tau)=\left(U^2\sigma \right)^k\mathcal{F}_{k}(1/U,V/U;1/\sigma,\tau/\sigma)\;.
\end{split}
\end{equation}
Using the crossing identities obeyed by the operator $\Upsilon$, it is then easy to find the crossing relations 
obeyed by the unconstrained function $\mathcal{H}_k$,
\begin{equation}\label{Hcrossing}
\mathcal{H}_{k}(U,V;\sigma,\tau)=\mathcal{H}_{k}(U/V,1/V;\tau,\sigma)=U^{2k}\sigma^{k-2} \mathcal{H}_{k}(1/U,V/U;1/\sigma,\tau/\sigma)\;.
\end{equation}

In closing, we should emphasize that the decomposition (\ref{scfwisol}) is not unique, since obviously one can add 
any  ``homogeneous'' term to ${\cal F}_k$ and subtract the same term from  ${\cal K}_k$. In the case of ${\cal N}=4$ super Yang-Mills,  where the solution of the superconformal Ward identity takes a similar form,  there is a  natural choice
for ${\cal F}_k$, namely the value of the correlator in the free field limit: ${\cal F}_k$ is then a simple rational function of $U$ and $V$. {\it A priori} there is no reason that an analogous natural choice  for   ${\cal F}_k$ should exist in the $(2, 0)$ theory,
but we will find experimentally that there is one, even in the absence (obvious) connection with free field theory.

\section{Position space}\label{pspace}
\subsection{The position space method}\label{pmethod}

In this subsection we describe a ``position-space method'' to compute one-half BPS correlators for the $AdS_7 \times S^4$ background,  analogous to the one used in \cite{Rastelli:2016nze,longads5} for the $AdS_5 \times S^5$ case.
This method mimics the standard recipe of evaluating holographic correlators as sums of Witten diagrams (see  \cite{longads5} for a quick review of the traditional method), but has the great advantage that no detailed knowledge of the supergravity effective action is required.

\begin{figure}[t]
 
\begin{subfigure}{0.5\textwidth}
\includegraphics[width=0.78\linewidth]{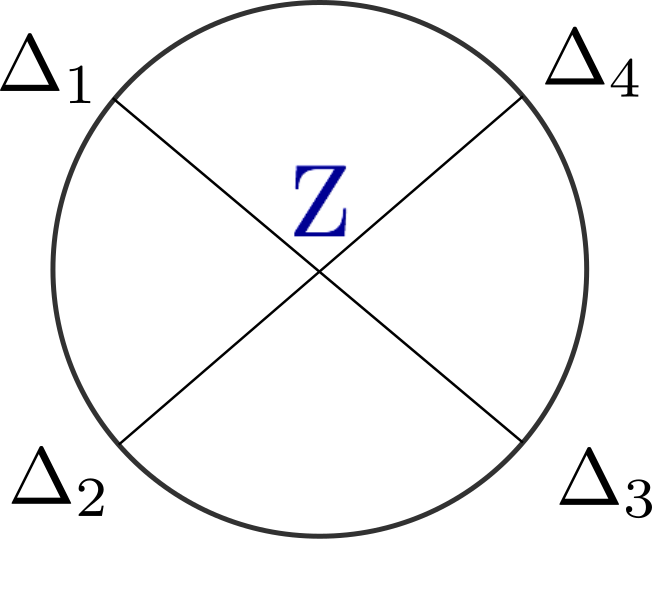} 
\caption{A contact Witten diagram}
\label{fcontact}
\end{subfigure}
\begin{subfigure}{0.5\textwidth}
\includegraphics[width=0.83\linewidth]{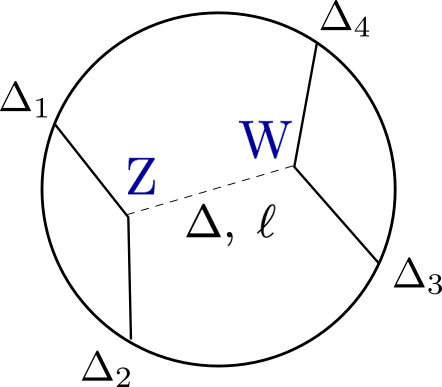}
\caption{An exchange Witten diagram}
\label{fexchange}
\end{subfigure}
\caption{Connected four-point Witten diagrams at tree level.}
\label{fWitten}
\end{figure}

By the standard AdS/CFT dictionary, the boundary correlator  $G_k$ is computed as a sum of bulk Witten diagrams. At tree level, the requisite Witten diagrams (Figure \ref{fWitten}) are either contact diagrams or exchange diagrams. They are evaluated
by assigning a bulk-to-boundary propagator $G_{B\partial}^{\Delta_i}$ to each external leg and a bulk-to-bulk propagator $G_{BB}^{\Delta}$ to the internal line. These propagators are Green's functions in AdS with appropriate boundary conditions. 
The simplest diagram is a contact Witten diagram without derivatives in the quartic vertex (Figure \ref{fcontact}). The  integral expression of such a diagram defines the so-called ``$D$-function'',
\begin{equation}
D_{\Delta_1\Delta_2\Delta_3\Delta_4}(\vec x_i)=\int_{AdS_{d+1}}dz\; G^{\Delta_1}_{B\partial}(z,\vec{x_1})\;G^{\Delta_2}_{B\partial}(z,\vec{x_2})\;G^{\Delta_3}_{B\partial}(z,\vec{x_3})\;G^{\Delta_4}_{B\partial}(z,\vec{x_4})\; ,
\end{equation}
where
\begin{equation}
G^{\Delta_i}_{B\partial}(z,\vec{x_i})=\left(\frac{z_0}{z_0^2+(\vec{z}-\vec{x}_i)^2}\right)^{\Delta_i}
\end{equation}
is the scalar bulk-to-boundary propagator \cite{Witten:1998qj}. Exchange  Witten diagrams (Figure \ref{fexchange}) are more complicated objects. However, when the twist ($\tau\colonequals \Delta-\ell$) of the exchanged field satisfies a special condition with respect to the external dimensions\footnote{For example in s-channel the condition is $\Delta_1+\Delta_2-\tau\in 2\mathbb{Z}_{\geqslant 0}$ and $\Delta_3+\Delta_4-\tau\in 2\mathbb{Z}_{\geqslant 0}$.}, a trick discovered in \cite{DHoker:1999aa} applies, and the exchange Witten diagrams can then be traded for a finite sum of  contact Witten diagrams. More details on this technique and an explanation of its mechanism in Mellin space are reviewed in \cite{longads5}.

While the traditional method requires the input of the precise effective Lagrangian, our position space method relies only on some qualitative features of the supergravity theory, to wit, the spectrum of supergravity fields, the selection rules encoded in the cubic vertices and a bound on the number of derivatives that can appear in  the quartic vertices. The spectrum of eleven dimensional supergravity compactified on $AdS_7\times S^4$ has long been worked out in \cite{vanNieuwenhuizen:1984iz} by diagonalizing the quadratic terms in the $AdS_7$ effective  action. The spectrum is organized in infinite Kaluza-Klein towers,  with increasing $USp(4)$ quantum numbers within each tower. Clearly the allowed exchanges between two pairs of half-BPS scalar KK modes are only a finite subset of the whole spectrum. There are two selection rules. The first selection rule is the R-symmetry selection rule which follows from elementary $USp(4)$ representation theory: the tensor product of two irreducible representations $[0,k]$ of $USp(4)$ yields a finite direct sum of irreps,
\begin{equation}
[0,k]\otimes[0,k]=\sum_{a=0}^k\sum_{b=0}^{a}[2(a-b),2b] \,.
\end{equation} 
 The fields with the admissible representations have been collected in Table \ref{ads7spectrum}. However not all fields {\it a priori} allowed by the R-symmetry selection rule will appear in the exchange channel. Additionally, there is another selection rule,
  which restricts the twist $\tau$ of a field that can couple to two scalars of dimensions $\Delta_1$ and $\Delta_2$,
  \begin{equation}
 \tau < \Delta_1+\Delta_2\,.  
 \end{equation}
 In the cases of interest here, $\Delta_1 = \Delta_2 = 2k$.
 The origin of this twist selection rule was discussed at length in \cite{longads5} and will not be repeated here. The spectrum and selection rules in the $AdS_7\times S^4$ supergravity  background are such that truncation conditions are satisfied and all exchange diagrams are given by finite sums of $D$-functions.

\begin{table}[t]
\begin{center}\begin{tabular}{cccccccccc}field & spin & $USp(4)$ irrep & $m^2$ & $\Delta$ & $k=2$ & $k=3$ & $k=4$ & $k=5$ & $k=6$ \\$\varphi_{\mu\nu,k}$ & 2 & $[0,k-2]$ & $4(k-2)(k+1)$ & $2k+2$ & {6} & 8 & {10} & 12 & {14} \\$A_{\mu,k}$ & 1 & $[2,k-2]$ & $4k(k-2)$ & $2k+1$ & {5} & 7 & {9} & 11 & {13} \\$C_{\mu,k}$ & 1 & $[2,k-4]$ & $4(k-1)(k+1)$ & $2k+3$ & - & - & {11} & 13 & {15} \\$s_{k}$ & 0 & $[0,k]$ & $4k(k-3)$ & $2k$ & {4} & 6 & {8} & 10 & {12} \\$t_{k}$ & 0 & $[0,k-4]$ & $4(k-1)(k+2)$ & $2k+4$ & - & - & {12} & 14 & 16 \\$r_{k}$ & 0 & $[4,k-4]$ & $4(k-2)(k+1)$ & $2k+2$ & - & - & {10} & 12 & {14}\end{tabular} \caption{\small KK modes contributing to exchange diagrams allowed by R-symmetry selection rules.}\label{ads7spectrum}
\end{center}
\end{table}

We also make an assumption on the quartic vertices. We assume that quartic vertices with at most  two spacetime derivatives effectively contribute to the four-point functions\footnote{The same assumption was  made in \cite{Rastelli:2016nze,longads5} for $AdS_5\times S^5$ based on the same reasoning, and was confirmed by explicit computation in \cite{Arutyunov:2017dti}.}. This is necessary in order to recover the correct flat-space limit 
as the radius of
 $AdS_7\times S^4$ is sent to infinity. A further technical  simplification ({\it c.f.} Appendix D of \cite{longads5}) is that when $\Delta_i\neq d=6$ the contribution from the zero-derivative vertices can be absorbed, by a redefinition of parameters, into the contribution of the two-derivative vertices. This will be useful  in concrete computations  to avoid redundancies in the parameterization of the quartic vertices. 

The position space method proceeds as follows. We use the above information on spectrum and vertices to write  a general ansatz for the supergravity amplitude which consists of an exchange part and a contact part,
\begin{equation}
\mathcal{A}_k(U,V;\sigma,\tau)=\mathcal{A}_{k,{\rm exchange}}(U,V;\sigma,\tau)+\mathcal{A}_{k,{\rm contact}}(U,V;\sigma,\tau)\;.
\end{equation}
Here we are working with the reduced correlator which depends only on the cross ratios and is obtained by stripping off the kinematic factor $\left(\frac{t_{12}\; t_{34}}{x_{12}^4x_{34}^4}\right)^k$. In this ansatz the exchange amplitude is summed over the three channels,  related to one another by crossing,
\begin{equation}
\mathcal{A}_{k,{\rm exchange}}(U,V;\sigma,\tau)=\mathcal{A}_{k, {\rm s-ex}}(U,V;\sigma,\tau)+\mathcal{A}_{k, {\rm t-ex}}(U,V;\sigma,\tau)+\mathcal{A}_{k, {\rm u-ex}}(U,V;\sigma,\tau)\;,
\end{equation}
\begin{equation}\label{Acrossing}
\begin{split}
\mathcal{A}_{k, {\rm t-ex}}(U,V;\sigma,\tau)={}&\left(\frac{U^2\tau}{V^2}\right)^k\mathcal{A}_{k,{\rm s-ex}}(V,U;\sigma/\tau,1/\tau)\;,\\
\mathcal{A}_{k, {\rm u-ex}}(U,V;\sigma,\tau)={}& \left(U^2\sigma \right)^k\mathcal{A}_{k,{\rm s-ex}}(1/U,V/U;1/\sigma,\tau/\sigma)\;.
\end{split}
\end{equation}
The s-channel exchange amplitude $\mathcal{A}_{k, {\rm s-ex}}$ is  given by the sum of all s-channel exchange Witten diagrams ompatible with the  selection rules for the cubic vertices. Schematically,
\begin{equation}
\mathcal{A}_{k, {\rm s-exchange}}=\sum_X \lambda_X Y_{R_X}(\sigma,\tau)\mathcal{E}_{X}(U,V)\;.
\end{equation}
Here  the label $X$ runs over the exchanged fields, $\mathcal{E}_X$ denotes the  corresponding exchange Witten diagram, $Y_{R_X}(\sigma,\tau)$ is the polynomial associated with the irreducible representation $R_X$ of the field $X$
 and finally $\lambda_X$ are the unknown coefficients to be determined. 
The formulae for evaluating the requisite exchange diagrams in terms contact diagrams have been given in Appendix C of \cite{longads5}, while a list of the relevant R-symmetry  polynomials is given in Appendix \ref{Rpoly}.
For the reader's convenience, we have also include these explicit expressions in the Mathematica notebook available from the ArXiv version of this paper. 

The discussion of contribution from  contact diagrams should distinguish two difference cases, as was explained in Appendix D of \cite{longads5}. When $\Delta_i=d$, we should include in the ansatz both the contribution from contact vertices with no derivatives and with two-derivatives. When $\Delta_i\neq d$, the zero-derivative contribution can be absorbed into the two-derivative contribution by redefining the parameters in the ansatz. Because of crossing symmetry, it is also convenient to split the two-derivative contribution into channels. The full contact vertex is the crossing symmetrization of the following s-channel contribution,
\begin{equation}
S_{\alpha_1\alpha_2\alpha_3\alpha_4}\int_{AdS_{d+1}}dX\; \frac{1}{16}(\triangledown s^{\alpha_1}\triangledown s^{\alpha_2}s^{\alpha_3}s^{\alpha_4}+s^{\alpha_3}s^{\alpha_4}\triangledown s^{\alpha_1}\triangledown s^{\alpha_2}) \,,
\end{equation}
where $\alpha_i$ collectively denote the indices of the symmetric traceless representations and $S_{\alpha_1\alpha_2\alpha_3\alpha_4}$ is an unspecified tensor symmetric under $(1\leftrightarrow2, 3\leftrightarrow4)$. When contracted with the R-symmetry null vectors the above vertices lead to the following contribution to the s-channel contact amplitude,
\begin{equation} \label{Akscont}
\mathcal{A}_{k,{\rm s-cont}}\propto\sum_{0\leq a+b\leq k}c_{ab}\sigma^a\tau^bU^{2k}(\bar{D}_{2k,2k,2k,2k}-\frac{2 \Gamma (2k )^2 \Gamma (4k -2)}{\Gamma (2k +1)^2 \Gamma (4k -3)}U\bar{D}_{2k,2k,2k+1,2k+1})\;.
\end{equation}
 Here we have used the so-called $\bar{D}$-functions\footnote{We emphasize that   $\bar{D}$-functions are independent of the spacetime dimension $d$. This is  clearest from their Mellin-Barnes representation,
\begin{equation}
\begin{split}
\bar{D}_{\Delta_1,\Delta_2,\Delta_3,\Delta_4}={}&\int \frac{ds}{2}\frac{dt}{2}U^{\frac{s}{2}}V^{\frac{t}{2}}\Gamma[-\frac{s}{2}]\Gamma[-\frac{s}{2}+\frac{\Delta_3+\Delta_4-\Delta_1-\Delta_2}{2}]\\
\times{}& \Gamma[-\frac{t}{2}]\Gamma[-\frac{t}{2}+\frac{\Delta_1+\Delta_4-\Delta_2-\Delta_3}{2}]\\
\times{}&\Gamma[\Delta_2+\frac{s+t}{2}]\Gamma[\frac{s+t}{2}+\frac{\Delta_1+\Delta_2+\Delta_3-\Delta_4}{2}]\, ,
\end{split}
\end{equation}
where $d$ completely drops out.
} defined by stripping off some kinematic factors from the $D$-functions,
\begin{equation}\label{dbar}
\frac{ \prod_{i=1}^4\Gamma(\Delta_i)}{\Gamma(\Sigma-\frac{1}{2}d)}\frac{2}{\pi^{\frac{d}{2}}}D_{\Delta_1\Delta_2\Delta_3\Delta_4}(x_1,x_2,x_3,x_4) \equalscolon \frac{r_{14}^{\Sigma-\Delta_1-\Delta_4}r_{34}^{\Sigma-\Delta_3-\Delta_4}}{r_{13}^{\Sigma-\Delta_4}r_{24}^{\Delta_2}}\bar{D}_{\Delta_1\Delta_2\Delta_3\Delta_4} (U,V)\;,
\end{equation}
with $\Sigma=\frac{1}{2}(\Delta_1+\Delta_2+\Delta_3+\Delta_4)$. The coefficients $c_{ab}$ in (\ref{Akscont}) are symmetric thanks to  the exchange symmetry  $(1\leftrightarrow2, 3\leftrightarrow4)$.
When $\Delta_i=d$, we need to also include  the zero-derivative contribution 
\begin{equation}
\sum_{0\leq a+b\leq k}c'_{ab}\sigma^a\tau^bU^{2k}\bar{D}_{2k,2k,2k,2k}\;,\quad\quad c'_{ab}=c'_{ba}\;.
\end{equation}
The crossed channel contributions $\mathcal{A}_{k,{\rm t-cont}}$ and $\mathcal{A}_{k,{\rm u-cont}}$ can then be obtained from $\mathcal{A}_{k,{\rm s-cont}}$ using the  crossing relation (\ref{Acrossing}).

Putting all these pieces together, we now have an anstaz $\mathcal{A}_k(U,V;\sigma,\tau)$ of the four-point function as a finite sum of $\bar{D}$-functions. It has polynomial dependence on $\sigma$ and $\tau$ and contains linearly all the unspecified coefficients $\lambda_X$, $c_{ab}$, $c'_{ab}$. These coefficients must to be fine-tuned in order to satisfy the superconformal Ward identity (\ref{scfwi}),
\begin{equation}
(\chi'\partial_{\chi'}-2\alpha'\partial_{\alpha'})\mathcal{G}_k(\chi,\chi';\alpha,\alpha')\big|_{\alpha'\to1/\chi'}=0\,.
\end{equation}
The ansatz $\mathcal{A}_k$ is not yet in a  form such  the superconformal Ward identity can be conveniently exploited. Fortunately, all $\bar{D}$-functions that appear in the ansatz  can be reached from the basic $\bar{D}$-function  $\bar{D}_{1111}$ with the repetitive use of six differential operators, 
\begin{equation}\label{Dbid}
\begin{split}
\bar{D}_{\Delta_1+1,\Delta_2+1,\Delta_3,\Delta_4}={}&\mathcal{D}_{12}\bar{D}_{\Delta_1,\Delta_2,\Delta_3,\Delta_4}:=-\partial_U \bar{D}_{\Delta_1,\Delta_2,\Delta_3,\Delta_4}\;,\\
\bar{D}_{\Delta_1,\Delta_2,\Delta_3+1,\Delta_4+1}={}&\mathcal{D}_{34}\bar{D}_{\Delta_1,\Delta_2,\Delta_3,\Delta_4}:=(\Delta_3+\Delta_4-\Sigma-U\partial_U )\bar{D}_{\Delta_1,\Delta_2,\Delta_3,\Delta_4}\;,\\
\bar{D}_{\Delta_1,\Delta_2+1,\Delta_3+1,\Delta_4}={}&\mathcal{D}_{23}\bar{D}_{\Delta_1,\Delta_2,\Delta_3,\Delta_4}:=-\partial_V \bar{D}_{\Delta_1,\Delta_2,\Delta_3,\Delta_4}\;,\\
\bar{D}_{\Delta_1+1,\Delta_2,\Delta_3,\Delta_4+1}={}&\mathcal{D}_{14}\bar{D}_{\Delta_1,\Delta_2,\Delta_3,\Delta_4}:=(\Delta_1+\Delta_4-\Sigma-V\partial_V )\bar{D}_{\Delta_1,\Delta_2,\Delta_3,\Delta_4}\;,\\
\bar{D}_{\Delta_1,\Delta_2+1,\Delta_3,\Delta_4+1}={}&\mathcal{D}_{24}\bar{D}_{\Delta_1,\Delta_2,\Delta_3,\Delta_4}:=(\Delta_2+U\partial_U+V\partial_V )\bar{D}_{\Delta_1,\Delta_2,\Delta_3,\Delta_4}\;,\\
\bar{D}_{\Delta_1+1,\Delta_2,\Delta_3+1,\Delta_4}={}&\mathcal{D}_{13}\bar{D}_{\Delta_1,\Delta_2,\Delta_3,\Delta_4}:=(\Sigma-\Delta_4+U\partial_U+V\partial_V )\bar{D}_{\Delta_1,\Delta_2,\Delta_3,\Delta_4}\;.
\end{split}
\end{equation}
The special function $\bar{D}_{1111}$ is in fact the familiar scalar one-loop box integral in four dimensions and will be denoted as  $\Phi$ from now on. It has a well-known representation in terms of dilogarithms,
\begin{equation}
\Phi(\chi,\chi')=\frac{1}{\chi-\chi'}\bigg(\log(\chi\chi')\log(\frac{1-\chi}{1-\chi'})+2{\rm Li}(\chi)-2{\rm Li}(\chi')\bigg)\;,
\end{equation}
and enjoys the following beautiful differential recursion relations \cite{Eden:2000bk}
\begin{equation}\label{recur}
\begin{split}
\partial_\chi\Phi={}&-\frac{1}{\chi-\chi'}\Phi-\frac{1}{\chi(\chi-\chi')}\ln(-1+\chi)(-1+\chi')+\frac{1}{(-1+\chi)(\chi-\chi')}\ln(\chi\chi')\;,\\
\partial_{\chi'}\Phi={}&\frac{1}{\chi-\chi'}\Phi+\frac{1}{\chi'(\chi-\chi')}\ln(-1+\chi)(-1+\chi')-\frac{1}{(-1+\chi')(\chi-\chi')}\ln(\chi\chi')\;.
\end{split}
\end{equation}
Using the above properties of $\bar{D}$-functions, we can unambiguously decompose the supergravity ansatz into a basis  spanned by $\Phi$, $\log U$, $\log V$ and 1,
\begin{equation}
\mathcal{A}_{k}(\chi,\chi';\alpha,\alpha')=R_\Phi \Phi(\chi,\chi')+R_{\log U}\log U+R_{\log V}\log V+R_1\, ,
\end{equation}
where the four coefficients  functions $R_\Phi$, $R_{\log U}$, $R_{\log V}$ and $R_1$ are rational functions of $\chi$, $\chi'$  and polynomials of $\alpha$, $\alpha'$. This decomposition  makes it straightforward to enforce the superconformal Ward identity (\ref{scfwi}) on $\mathcal{A}_k$. Upon acting on $\mathcal{A}_{k}(\chi,\chi';\alpha,\alpha')$ with the differential operator $(\chi'\partial_{\chi'}-2\alpha'\partial_{\alpha'})$ from (\ref{scfwi}) and setting $\alpha'=1/\chi'$, a new set of coefficient functions $\widetilde{R}_\Phi$, $\widetilde{R}_{\log U}$, $\widetilde{R}_{\log V}$ $\widetilde{R}_1$ are generated from $R_\Phi$, $R_{\log U}$, $R_{\log V}$, $R_1$ with the help of the differential recursion relation of $\Phi$.  The superconformal Ward identity then dictates the following conditions
\begin{equation}\label{scfwicoefunc}
\begin{split}
{}&\widetilde{R}_\Phi(\chi,\chi';\alpha,1/\chi')=0\;,\\
{}&\widetilde{R}_{\log U}(\chi,\chi';\alpha,1/\chi')=0\;,\\
{}&\widetilde{R}_{\log V}(\chi,\chi';\alpha,1/\chi')=0\;,\\
{}&\widetilde{R}_1(\chi,\chi';\alpha,1/\chi')=0\;,
\end{split}
\end{equation}
which imply a large set of linear equations for the unknown coefficients. This set of equations is constraining enough to fix all relative coefficients up to an overall constant. That the overall constant should remain undetermined is inevitable because the condition (\ref{scfwi}) is homogeneous. To fix it, we demand that the OPE coefficient of the intermediate one-half BPS operator ${\cal O}^{(2)}$ has the correct value.  
 The details of this calculation are discussed in Appendix \ref{lastcoe}.

\subsection{The $k=2$ four-point function}
In the rest of this section we will put the above position method to work. We start with the $k=2$ one-half BPS operator $O^{(2)}$ which sits in the same short supermultiplet as the stress tensor. Its four-point function $G_2$ was first calculated in \cite{Arutyunov:2002ff} and we will reproduce their result. By the two selection rules of cubic vertices the allowed exchanges are identified to be all the fields that belong to the $k=2$ family in Table \ref{ads7spectrum}. Explicitly, the exchange Witten diagrams in the s-channel contribute
\begin{equation}
\mathcal{A}_{2,\; {\rm s-exchange}}=Y_{11}\lambda_{s_2}\mathcal{E}_{s_2}+Y_{10}\lambda_{A_2}\mathcal{E}_{A_2}+\lambda_{\varphi_2}\mathcal{E}_{\varphi_2}\;.
\end{equation}
As was discussed above, the contribution of contact Witten diagrams can be split into channels and then cross-symmetrized. Moreover, because $\Delta_i\neq d$ we can absorb the contribution of the zero-derivative terms into the two-derivative terms. Hence we have the following s-channel ansatz for the contact contributions,
\begin{equation}
\mathcal{A}_{2,\; {\rm s-contact}}=\sum_{0\leq a+b\leq 2}c_{ab}\tau^a\sigma^b\frac{\pi^3 U^4}{432}(8\bar{D}_{4444}-5U\bar{D}_{5544})
\end{equation}
 where $c_{ab}=c_{ba}$ follows from  symmetry under exchanging  operators 1 and 2. The total amplitude $\mathcal{A}_{2}$ is  obtained from cross-symmetrizing the above s-channel amplitude,
\begin{equation}
\begin{split}
\mathcal{A}_{2}(U,V;\sigma,\tau)={}&\mathcal{A}_{2,s}(U,V;\sigma,\tau)+\left(\frac{U^2\tau}{V^2}\right)^2\mathcal{A}_{2,s}(V,U;\sigma/\tau,1/\tau)\\
{}&+\left(U^2\sigma \right)^2\mathcal{A}_{2,s}(1/U,V/U;1/\sigma,\tau/\sigma)\;,\\
\mathcal{A}_{2,s}(U,V;\sigma,\tau)={}&\mathcal{A}_{2,\;{\rm s-exchange}}(U,V;\sigma,\tau)+\mathcal{A}_{2,\;{\rm s-contact}}(U,V;\sigma,\tau)\;.
\end{split}
\end{equation} 
Decomposing this ansatz into the basis of functions $\Phi$, $\log U$, $\log V$ and 1 and enforcing the superconformal Ward identity (\ref{scfwicoefunc}), we find enough constraints to fix all the coefficients up to an overall factor $\xi$,
\begin{equation}\label{keq2a}
\begin{split}
{}& \lambda_{s_2}=\xi,\;\;\;\lambda_{A_2}=-\frac{1}{9}\xi,\;\;\;\lambda_{\varphi_2}=\frac{1}{576}\xi,\\
{}& c_{00}=\frac{1}{36}\xi,\;\;\;c_{01}=-\frac{1}{9}\xi,\;\;\;c_{02}=\frac{1}{36}\xi,\;\;\;c_{11}=-\frac{1}{12}\xi\;.
\end{split}
\end{equation}
The last coefficient can be determined by demanding that the relevant term in the OPE  is compatible with the known value of the three-point coupling $\langle {\cal O}^2  {\cal O}^2  {\cal O}^2 \rangle$. The details of this computation are relegated to Appendix \ref{lastcoe}). The  result is
\begin{equation}\label{keq2b}
\xi=\frac{15552}{\pi ^3 n^3}\;.
\end{equation}

By setting the R-symmetry cross-ratios to the special value $\alpha=\alpha'=1/\chi'$ we find the following ``holomorphic'' correlator,
\begin{equation}\label{keq2holo}
\mathcal{A}_2(\chi,\chi';1/\chi',1/\chi')=\frac{2 \chi ^2 ((\chi -1) \chi +1)}{n^3 (\chi -1)^2}\;.
\end{equation}
As we show in Appendix \ref{4ptchiral}, it reproduces the corresponding $\mathcal{W}_\infty$ four-point function.

The full four-point function  can also be massaged into a form consistent with the solution to the superconformal Ward identity\footnote{In principle this can be done solely using   $\bar{D}$-functions identities as in \cite{Dolan:2001tt,Arutyunov:2002ff}, but 
in practice it more efficient to guess and then check. The guesswork starts with the easier step of obtaining $\mathcal{H}_k$ from  $\widetilde{\mathcal{M}}_k$ (see below in Section \ref{manifestscf}). We then check this guess by decomposing into the basis spanned by $\Phi$, $\log U$, $\log V$ and 1: the difference $\mathcal{G}-\Upsilon\circ \mathcal{H}$ should give a rational function.
}. We found that the result can be written as
\begin{equation}\label{keq2reduced}
\mathcal{A}_2=\frac{\sigma  \tau  U^4}{n^3 V^2}+\frac{\sigma  U^2}{n^3}+\frac{\tau  U^2}{n^3 V^2}+\Upsilon \circ\left(\frac{U^5}{2n^3V}\bar{D}_{7333}\right)\;
\end{equation}
where the differential operator $\Upsilon$ was defined in (\ref{upsilon}). This agrees with the result obtained by Arutynov and Sokatchev \cite{Arutyunov:2002ff}.
Remarkably, the inhomogeneous term ${\cal F}_2$ is a rational function of the cross ratios, just as in ${\cal N}=4$ super-Yang Mills, where the analogous inhomogeneous term can be identified with the free-field limit of the correlator.
Of course, one could always {\it choose} the inhomogeneous term ${\cal F}$ to be  rational, at the expense of making the dynamical function ${\cal H}$ more complicated.\footnote{The only requirement on ${\cal F}$ is that upon setting $\alpha = \alpha' = 1/\chi'$
it should reduce to the chiral algebra correlator. It is always possible to find a rational function with this property.} The non-trivial statement is that we can cast the correlator into a form where ${\cal F}$ is rational and ${\cal H}$ is a linear combination of $\bar D$-functions.
This property will persist in the $k=3$ and $k=4$ examples that we study below. We conjecture that it holds in general.

\subsection{The $k=3$ four-point function}\label{keq3}
We now proceed to the next  Kaluza-Klein level. The allowed exchanges for $G_3$ include the three component fields of the $k=2$ family in Table \ref{ads7spectrum} and all other fields of the $k=4$  family  except for the field $t_4$. This field is ruled out because it has twist 12, which  
violates the twist upper bound. The $k=3$ family, on the other hand, is absent because of the R-symmetry selection rule. For the contact diagrams  we notice that in this case the conformal dimension of the external operators coincides with the boundary spacetime dimension $\Delta=d$. As was discussed in the Appendix D of \cite{longads5}, the zero-derivative contribution can no longer be  reabsorbed into the two-derivative contribution. We need to include in our ansatz both set of parameters for the quartic vertices, even if this will lead to some (harmless) ambiguities in fixing the coefficients of contact vertices when we use the superconformal Ward identity.

The s-channel ansatz is again given by an exchange part $\mathcal{A}_{3,\; {\rm s-exchange}}$ and a contact part $\mathcal{A}_{3, {\rm s-contact}}$
\begin{equation}
\begin{split}
\mathcal{A}_{3,\; {\rm s-exchange}}=&Y_{11}\lambda_{s_2}\mathcal{E}_{s_2}+Y_{10}\lambda_{A_2}\mathcal{E}_{A_2}+Y_{00}\lambda_{\varphi_2}\mathcal{E}_{\varphi_2}\\
+& Y_{22}\lambda_{s_4}\mathcal{E}_{s_4}+Y_{21}\lambda_{A_4}\mathcal{E}_{A_4}+Y_{11}\lambda_{\varphi_4}\mathcal{E}_{\varphi_4}+Y_{20}\lambda_{r_4}\mathcal{E}_{r_4}+Y_{10}\lambda_{C_4}\mathcal{E}_{C_4}\;,
\end{split}
\end{equation}
\begin{equation}
\mathcal{A}_{3, {\rm s-contact}}=\sum_{0\leq a+b\leq 3}c_{ab}\tau^a\sigma^b\frac{7\pi^3 U^6}{72000}(2\bar{D}_{6666}-U\bar{D}_{7766})+\sum_{0\leq a+b\leq 3}c'_{ab}\tau^a\sigma^b\frac{7\pi^3 U^6}{36000}\bar{D}_{6666}
\end{equation}
where the coefficients $c_{ab}=c_{ba}$, $c'_{ab}=c'_{ba}$ are symmetric. The total amplitude ansatz is obtained by further including the t-channel and u-channel contributions which are obtained from the above s-channel contribution by crossing. Due to the ambiguity in the parameterizing of contact vertices, the superconformal Ward identity fixes only the coefficients of the exchange diagrams, leaving a subset of $c_{ab}$, $c'_{ab}$ unfixed,
\begin{equation}\label{keq3a}
\begin{split}
{}& \lambda_{s_2}=\xi,\;\;\;\lambda_{A_2}=-\frac{3}{50}\xi,\;\;\;\lambda_{\varphi_2}=\frac{1}{3600}\xi,\\
{}& \lambda_{s_4}=\frac{224}{135}\xi,\;\;\;\lambda_{A_4}=-\frac{8}{105}\xi,\;\;\;\lambda_{\varphi_4}=\frac{1}{5040}\xi,\;\;\; \lambda_{C_4}=-\frac{1}{4900}\xi,\;\;\; \lambda_{r_4}=\frac{16}{945}\xi,\\
{}& c_{12}=\frac{1}{140}\xi,\;\;\; c'_{03}=\frac{1}{630}\xi,\;\;\; c'_{11}=-\frac{17}{630}\xi,\;\;\; c'_{12}=-\frac{13}{1260}\xi\;. 
\end{split}
\end{equation}
These unfixed coefficients are actually redundant: the corresponding expressions are proportional to a sum of $\bar{D}$-functions which is zero in disguise,  thanks to $\bar{D}$-function identities. We can then  set them  to zero (or to any convenient value). 
In Appendix \ref{lastcoe} we fix the last coefficient by enforcing the correct value of the OPE coefficient of ${\cal O}^{(2)}$,
which gives
\begin{equation}\label{keq3b}
\xi= \frac{1080000}{n^3\pi^3}\;.
\end{equation}
By setting $\alpha=\alpha'=1/\chi'$, we extract another ``holomorphic'' correlator
\begin{equation}\label{keq3holo}
\mathcal{A}_3(\chi,\chi';1/\chi',1/\chi')=\frac{9 \chi ^2 \left(2 \chi ^6-6 \chi ^5+9 \chi ^4-8 \chi ^3+9 \chi ^2-6 \chi +2\right)}{4 n^3 (\chi -1)^4}\; ,
\end{equation}
which agrees with the expected result from $W_\infty$ (see Appendix \ref{4ptchiral}).
As was the case for $k=2$, the $k=3$ supergravity four-point function can also be put into a compact form that manifestly solves the superconformal Ward identity,
\begin{equation}
\mathcal{A}_{3}=\mathcal{F}_{3,{\rm conn}}+\Upsilon\circ\mathcal{H}_{3}\;.
\end{equation}
Here $\mathcal{F}_{3,{\rm conn}}$ is a simple rational function of the cross ratios, 
\begin{equation}\label{keq3reduceda}
\mathcal{F}_{3,{\rm conn}}=\frac{9}{4n^3}\left(\frac{\sigma  \tau ^2 U^6}{V^4}+\frac{\sigma ^2 \tau  U^6}{V^2}+\sigma ^2 U^4+\frac{\tau ^2 U^4}{V^4}+\sigma  U^2+\frac{\tau  U^2}{V^2}\right)\;,
\end{equation}
while $\mathcal{H}_{3}$ is  given in terms of $\bar{D}$-functions by the following expression,
\begin{equation}\label{keq3reducedb}
\begin{split}
\mathcal{H}_{3}={}&\frac{U^5}{48 n^3 V}\bigg(U^2 (9 \bar{D}_{9533}+7 \bar{D}_{9544}+2 \bar{D}_{9555})+\sigma  (9 \bar{D}_{3539}+7 \bar{D}_{4549}+2 \bar{D}_{5559})\\
{}&+\tau  V^2 (9 \bar{D}_{3593}+7 \bar{D}_{4594}+2 \bar{D}_{5595})\bigg)\;.
\end{split}
\end{equation}

\subsection{The $k=4$ four-point function}\label{keq4}
The calculation of the $G_4$ correlator is very similar to the calculation of $G_2$. The twist and R-symmetry selection rules dictate that the allowed exchange fields are all the component fields from the $k=2,3,4$ families in Table \ref{ads7spectrum}, except for $t_6$. This leads to the following s-channel exchange ansatz,
\begin{equation}
\begin{split}
\mathcal{A}_{4,\; {\rm s-exchange}}=&Y_{11}\lambda_{s_2}\mathcal{E}_{s_2}+Y_{10}\lambda_{A_2}\mathcal{E}_{A_2}+Y_{00}\lambda_{\varphi_2}\mathcal{E}_{\varphi_2}\\
+& Y_{22}\lambda_{s_4}\mathcal{E}_{s_4}+Y_{21}\lambda_{A_4}\mathcal{E}_{A_4}+Y_{11}\lambda_{\varphi_4}\mathcal{E}_{\varphi_4}+Y_{20}\lambda_{r_4}\mathcal{E}_{r_4}+Y_{10}\lambda_{C_4}\mathcal{E}_{C_4}+Y_{00}\lambda_{t_4}\mathcal{E}_{t_4}\\ 
+&Y_{33}\lambda_{s_6}\mathcal{E}_{s_6}+Y_{32}\lambda_{A_6}\mathcal{E}_{A_6}+Y_{22}\lambda_{\varphi_6}\mathcal{E}_{\varphi_6}+Y_{31}\lambda_{r_6}\mathcal{E}_{r_6}+Y_{21}\lambda_{C_6}\mathcal{E}_{C_6}\;.
\end{split}
\end{equation}
Furthermore, the contribution of the zero-derivative contact vertices can be absorbed into the parameterization of the two-derivative contribution. The ansatz for the contact diagrams is thus
\begin{equation}
\mathcal{A}_{4,\; {\rm s-contact}}=\sum_{0\leq a+b\leq 4}c_{ab}\tau^a\sigma^b\frac{11\pi^3 U^8}{474163200}(32\bar{D}_{8888}-13U\bar{D}_{9988})
\end{equation}
with $c_{ab}=c_{ba}$.

 Imposing the superconformal Ward identity determines all the coefficients in terms of a single parameter $\xi$,
\begin{equation}
\begin{split}\label{keq4a}
{}& \lambda_{s_2}=\xi,\;\;\;\lambda_{A_2}=-\frac{2}{49}\xi,\;\;\;\lambda_{\varphi_2}=\frac{1}{12544}\xi,\\
{}&  \lambda_{s_4}=\frac{125}{42}\xi,\;\;\;\lambda_{A_4}=-\frac{125}{1134}\xi,\;\;\;\lambda_{\varphi_4}=\frac{5}{36288}\xi,\;\; \lambda_{C_4}=-\frac{1}{588}\xi,\;\; \lambda_{r_4}=\frac{125}{756}\xi,\;\; \lambda_{t_4}=\frac{1}{1960}\xi,\\
{}&  \lambda_{s_6}=\frac{33}{28}\xi,\;\;\;\lambda_{A_4}=-\frac{5}{154}\xi,\;\;\;\lambda_{\varphi_4}=\frac{5}{177408}\xi,\;\; \lambda_{C_4}=-\frac{25}{548856}\xi,\;\; \lambda_{r_4}=\frac{1}{154}\xi,\\
{}& c_{00}=-\frac{5}{1848}\xi,\;\;\; c_{01}=-\frac{5}{1386}\xi,\;\;\;c_{02}=\frac{155}{2772}\xi,\;\;\;c_{03}=-\frac{5}{1386}\xi,\;\;\;c_{04}=-\frac{5}{1848}\xi,\;\;\;\\
{}& c_{11}=\frac{1205}{5544}\xi,\;\;\; c_{12}=\frac{1235}{5544}\xi,\;\;\;c_{13}=-\frac{5}{2772}\xi,\;\;\;c_{22}=\frac{25}{396}\xi\,.
\end{split}
\end{equation}
Finally, $\xi$ is fixed by comparing with the OPE coefficient of the intermediate ${\cal O}^{(2)}$ operator (see Appendix  \ref{lastcoe}),
\begin{equation}\label{keq4b}
\xi=\frac{16595712}{n^3\pi^3}\;.
\end{equation}
The ``holomorphic'' correlator is given by
\begin{equation}\label{keq4holo}
\mathcal{A}_4(\chi,\chi';1/\chi',1/\chi')=\frac{8 \chi ^2 \left(\chi ^2-\chi +1\right)^2 \left(5 \chi ^6-15 \chi ^5+9 \chi ^4+7 \chi ^3+9 \chi ^2-15 \chi +5\right)}{5 n^3 (\chi -1)^6}\; ,
\end{equation}
which is again in agreement with the $W_\infty$ result, see Appendix \ref{4ptchiral}.
Finally, we recast the result  in a way that makes superconformal symmetry manifest,
\begin{equation}
\mathcal{A}_{4}=\mathcal{F}_{4,{\rm conn}}+\Upsilon\circ\mathcal{H}_{4}\;.
\end{equation}
Here,  the ``inhomogenous'' term $\mathcal{F}_{4,{\rm conn}}$ is the following rational function of the cross ratios,
\begin{equation}\label{keq4reduceda}
\begin{split}
\mathcal{F}_{4,{\rm conn}}={}&\frac{4 U^2 \left(\sigma  \tau ^3 U^6+\sigma ^3 \tau  U^6 V^4+\tau ^3 U^4+\sigma ^3 U^4 V^6+\sigma  V^6+\tau  V^4\right)}{n^3 V^6}\\
{}&+\frac{718 U^4 \left(\tau ^2+\sigma ^2 \tau ^2 U^4+\sigma ^2 V^4\right)}{275 n^3 V^4}-\frac{776 \sigma  \tau  U^4 \left(\tau  U^2+\sigma  U^2 V^2+V^2\right)}{275 n^3 V^4}\;,
\end{split}
\end{equation}
while the dynamical  function $\mathcal{H}_4$ is a linear combination of $\bar{D}$-functions. It is most conveniently represented as a sum over three channels,
\begin{equation}\label{keq4reducedb}
\mathcal{H}_4=\mathcal{H}_{4s}+\mathcal{H}_{4t}+\mathcal{H}_{4u}\;.
\end{equation}
The s-channel is given by
\begin{equation}\label{keq4reducedc}
\begin{split}
\mathcal{H}_{4s}={}&\frac{U^9}{475200 n^3 V}\bigg(7920 \bar{D}_{11,7,3,3}+8496 \bar{D}_{11,7,4,4}+4040 \bar{D}_{11,7,5,5}\\
{}&+1100 \bar{D}_{11,7,6,6}+165 \bar{D}_{11,7,7,7}\bigg)+\frac{\tau  U^9}{118800 n^3 V^3}\bigg(4308 \bar{D}_{11,5,3,5}\\
{}&+825 V^2 \bar{D}_{11,7,5,5}+605 V^2 \bar{D}_{11,7,6,6}+165 V^2 \bar{D}_{11,7,7,7}+3088 V \bar{D}_{11,6,4,5}\\
{}&+2240 V \bar{D}_{11,6,5,6}+605 V \bar{D}_{11,6,6,7}+3088 \bar{D}_{11,5,4,6}+825 \bar{D}_{11,5,5,7}\bigg)\, ,
\end{split}
\end{equation}
and the t- and u-channels are obtained  by the substitution rules
\begin{equation}\label{keqreducedd}
\begin{split}
\mathcal{H}_{4t}={}&\mathcal{H}_{4s}(U\to U/V,V\to 1/V;\sigma\to\tau,\tau\to\sigma)\;,\\
\mathcal{H}_{4u}={}&U^8\sigma^2\mathcal{H}_{4s}(U\to 1/U,V\to V/U;\sigma\to1/\sigma,\tau\to\tau/\sigma)\;.
\end{split}
\end{equation}

\section{Mellin space}\label{mmethod}
The position space method discussed in the previous section is a rigorous algorithm to compute four-point functions without {\it a priori} knowing the vertices. Though much easier than the traditional approach,  this method also runs 
out of steam as one moves up in the Kaluza-Klein towers due to unmanageable computational complexity and does not appear promising to provide a general solution. What's more, the position space method presents the final answer as an unintuitive sum of contact Witten diagrams.  The most suitable language to address both problems is the Mellin representation initiated by Mack \cite{Mack:2009mi} and developed by Penedones and others \cite{Penedones:2010ue,Paulos:2011ie,Fitzpatrick:2011ia,Costa:2012cb,Fitzpatrick:2012cg,Costa:2014kfa,Goncalves:2014rfa}.\footnote{For other applications and recent developments, see \cite{Paulos:2012nu,Nandan:2013ip,Lowe:2016ucg,Rastelli:2016nze,Paulos:2016fap,Nizami:2016jgt,Gopakumar:2016cpb,Gopakumar:2016wkt,Dey:2016mcs,Aharony:2016dwx, Rastelli:2017ecj,Dey:2017fab,Dey:2017oim}). } We refer the reader to \cite{longads5} for a detailed review of this formalism. We will be very brief here.

Using the inverse Mellin transformation, we express the connected part of the four-point function as a double integral
\begin{equation}\label{mellinG}
\begin{split}
G_{k,{\rm conn}}={}&\sum_{L+M+N=k}a^Lb^Mc^N\int_{-i\infty}^{i\infty} \frac{ds}{2}\; \frac{dt}{2}\; A^{s/2-2k}B^{u/2-2k}C^{t/2-2k}\\
{}&\times \mathcal{M}_{k,\;LMN}(s,t)\Gamma^2[-\frac{s}{2}+2k]\Gamma^2[-\frac{t}{2}+2k]\Gamma^2[-\frac{u}{2}+2k]\;.
\end{split}
\end{equation}
For notational convenience we have introduced 
\begin{equation}
\begin{split}
A \colonequals {}&x^2_{12}x^2_{34}\;,\;\;\;\;B \colonequals x^2_{13}x^2_{24}\;,\;\;\;\;C \colonequals x^2_{14}x^2_{23}\;,\\
a \colonequals{}&t_{12}t_{34}\;,\;\;\;\;b \colonequals t_{13}t_{24}\;,\;\;\;\;c \colonequals t_{14}t_{23}\;.\\
\end{split}
\end{equation}
The variables $s$, $t$ and $u$ are interpreted as  ``Mandelstam variables''\footnote{There exist many different definitions for $s$, $t$ and $u$ in the Mellin amplitude literature. We use here the convention of \cite{Rastelli:2016nze,longads5} which is the most natural definition from the analogy with flat space scattering amplitude.} and satisfy the constraint $s+t+u=8k$. The integration of the two independent variables $s$, $t$ is performed  along the imaginary axis. We further define  the {\it Mellin amplitude} $\mathcal{M}(s,t;\sigma,\tau)$ a sum of the partial amplitudes from $\mathcal{M}_{k,\;LMN}$,
\begin{equation}
\mathcal{M}(s,t;\sigma,\tau)=a^{-k}\sum_{L+M+N=k}a^Lb^Mc^N\mathcal{M}_{k,\;LMN}(s,t)\;.
\end{equation}
The above representation (\ref{mellinG}) of the crossing invariant correlator $G_{k,{\rm conn}}$ makes the following symmetry properties of the partial amplitudes clear,
\begin{equation}
\begin{split}
\mathcal{M}_{k,\;NML}(t,s)={}&\mathcal{M}_{k,\;LMN}(s,t)\;,\\
\mathcal{M}_{k,\;MNL}(u,t)={}&\mathcal{M}_{k,\;LMN}(s,t)\;,
\end{split}
\end{equation}
which in turn imply  simple crossing symmetry relations for the Mellin amplitude $\mathcal{M}$,
\begin{equation}
\begin{split}
\sigma^{k} {\mathcal{M}}_k( u,t;1/\sigma,\tau/\sigma)={}& {\mathcal{M}}_k(s,t,;\sigma,\tau)\;,\\
\tau^{k} {\mathcal{M}}_k(t,s;\sigma/\tau,1/\tau)={}& {\mathcal{M}}_k(s,t;\sigma,\tau)\;.
\end{split}
\end{equation}

The Mellin amplitude bears several formal  similarities  with the  flat space  S-matrix. Being analogous to amputated on-shell tree-level Feynman diagrams in flat space, tree-level Witten diagrams admit extremely simple representations in Mellin space.  If we temporarily suppress the dependence on R-symmetry variables, then (\ref{mellinG}) reduces to 
\begin{equation} \label{GkM}
G_{k, {\rm conn}}=\int_{-i\infty}^{i\infty} \frac{ds}{2}\; \frac{dt}{2}\; A^{s/2-2k}B^{u/2-2k}C^{t/2-2k}\mathcal{M}(s,t)\Gamma^2[-\frac{s}{2}+2k]\Gamma^2[-\frac{t}{2}+2k]\Gamma^2[-\frac{u}{2}+2k]\;.
\end{equation}
The contact Witten diagram with a nonderivative vertex is just a constant in Mellin space. More generally, the contact Witten diagram with a $2n$-derivatives vertex yields a polynomial of degree $n$ in the Mandelstam variables \cite{Penedones:2010ue}. Exchange diagrams also have a simple structure in Mellin space, as a sum over simple poles with polynomial residues plus another regular polynomial piece \cite{Costa:2012cb},
\begin{equation}\label{mellinexchnage}
\mathcal{M}_{\rm exchange}(s,t)= \sum_{m=0}^{\infty}\frac{Q_{J,m}(t)}{s-\tau-2m}+P_{J-1}(s,t)
\end{equation}
Here we have considered an s-channel exchange Witten diagram with an exchanged field of conformal dimension $\Delta$ and spin $J$ (and hence twist $\tau=\Delta-J$). The functions $Q_{J,m}(t)$ and $P_{J-1}(s,t)$ 
are polynomials of  degrees $J$ and $J-1$, respectively. 
When the twist of the exchanged field satisfies the condition
\begin{equation}\label{spectrcond}
\tau=4k-2m\;, \quad\quad m\in \mathbb{Z}_{\geqslant 0}
\end{equation}
a remarkable simplification occurs: the infinite sum in (\ref{mellinexchnage}) truncates to finitely many terms (see Section 3.2 of \cite{longads5} for a detailed discussion). This is the Mellin-space counterpart of the fact that certain exchange Witten diagrams can be evaluated as finitely many contact Witten diagrams. The spectrum condition (\ref{spectrcond}) is precisely met by the eleven dimensional supergravity on $AdS_7\times S^4$.  For this reason the analytic structure of the Mellin amplitude of the $AdS_7\times S^4$ four-point function is extremely simple: it has just finitely many simple poles and some extra polynomial terms.

\subsection{Superconformal symmetry in Mellin space}\label{manifestscf}
In position space, superconformal symmetry of the four-point function is encoded in the solution of the superconformal Ward identity,\footnote{As the Mellin transformation of the disconnected part of the correlator is ill-defined, we focus on the connected part.}  
\begin{equation} \label{wardagain}
{\cal G}_{k, {\rm conn}} = {\cal F}_{k, {\rm conn}} + \Upsilon \circ {\cal H}_{k, {\rm conn}} \, .
\end{equation}
 Our position space results have led us to
 conjecture  that  in the supergravity limit we can take  ${\cal H}_k$ to be a linear combination of $\bar D$-functions, and
 $\mathcal{F}_{k,{\rm conn}}$ to be rational function of the cross ratios.  The details of this section are very technical, but the main result is easy to summarize. We are going to find that Mellin version of (\ref{wardagain}) takes the form
 \begin{equation}
 \mathcal{M}_k=\widehat{\Theta}\circ\widetilde{\mathcal{M}}_k\; ,
 \end{equation}
 where $\widetilde{\mathcal{M}}_k$ is an ``auxiliary'' Mellin amplitude defined from the dynamical function  ${\cal H}_{k, {\rm conn}}$, see (\ref{mellinH}), and $\widehat{\Theta}$ a somewhat complicated difference operator defined in (\ref{Thete}).

We start by writing the Mellin  transformation of the left-hand side of (\ref{wardagain}) using (\ref{mellinG}),
\begin{equation}
\mathcal{G}_{k,{\rm conn}}=\int_{-i\infty}^{i\infty} \frac{ds}{2}\frac{dt}{2} U^{\frac{s}{2}}V^{\frac{t}{2}-2k} \mathcal{M}_k(s,t;\sigma,\tau)\Gamma^2[-\frac{s}{2}+2k]\Gamma^2[-\frac{t}{2}+2k]\Gamma^2[-\frac{u}{2}+2k]\;.
\end{equation}
Consider now the right-hand side of (\ref{wardagain}).
The Mellin transform of  the rational function $\mathcal{F}_{k,{\rm conn}}$ is ill-defined. As explained in  \cite{longads5}, it can be defined as ``zero''.\footnote{$\mathcal{F}_{k,{\rm conn}}$ 
 can be recovered as a subtle regularization effect in properly defining the contour integrals of the inverse Mellin transformation \cite{longads5}.}
On the other hand, we can write a
Mellin representation for the dynamical function $\mathcal{H}_k$ in terms of an ``auxiliary'' Mellin amplitude $\widetilde{\mathcal{M}}_k(s,t;\sigma,\tau)$
\begin{equation}\label{mellinH}
\mathcal{H}_k=\int_{-i\infty}^{i\infty} \frac{ds}{2}\frac{dt}{2} U^{\frac{s}{2}+1}V^{\frac{t}{2}-2k+1} \widetilde{\mathcal{M}}_k(s,t;\sigma,\tau)\Gamma^2[-\frac{s}{2}+2k]\Gamma^2[-\frac{t}{2}+2k]\Gamma^2[-\frac{\tilde{u}}{2}+2k]
\end{equation}
where $\tilde{u}=u-6$. As we demonstrate the shift has the virtue of giving simple transformation properties to $\widetilde{\mathcal{M}}_k(s,t;\sigma,\tau)$  under crossing,
\begin{equation}\label{crossingmtilde}
\begin{split}
\nonumber \sigma^{k-2} {\widetilde {\mathcal M}}_k(\tilde u,t;1/\sigma,\tau/\sigma)={}& {  \widetilde {\mathcal M}}_k(s,t,;\sigma,\tau)\;,\\
\tau^{k-2} {  \widetilde {\mathcal M}}_k(t,s;\sigma/\tau,1/\tau)={}& { \widetilde {\mathcal M}}_k(s,t;\sigma,\tau)\;.
\end{split}
\end{equation}
In the auxiliary amplitude $\widetilde{\mathcal{M}}_k(s,t;\sigma,\tau)$, the triplet variables $(s,t,\tilde{u})$ replaces $(s,t,u)$ to become the set of variables that permute under crossing. This becomes especially evident after we restore all the factors of $x_{ij}^2$ and $t_{ij}$. Restoring this factor will also facilitate the extraction of the difference operator.
 Let us see this in detail.

We first note that the combination 
\begin{equation} \label{insertee}
a^kA^{-2k}\Upsilon\circ\mathcal{H}_k
\end{equation}
 is crossing invariant, since it has the same crossing properties as $G_k$. Upon inserting the inverse Mellin transformation (\ref{mellinH}) into (\ref{insertee})
  and decomposing the auxiliary amplitude with respect to the R-symmetry monomials,
\begin{equation}
\widetilde{\mathcal{M}}(s,t;\sigma,\tau)=\sum_{l+m+n=k-2}\sigma^m\tau^n\;\widetilde{\mathcal{M}}_{k,\;lmn}(s,t)\;,
\end{equation}
we find
\begin{equation}
(\frac{a}{A^2})^k\;\Upsilon\circ\sum_{l+m+n=k-2}\int_{-i\infty}^{i\infty} \frac{ds}{2}\; \frac{dt}{2}\; U^{s/2+1} V^{t/2-2k+1} \sigma^m \tau^n \widetilde{\mathcal{M}}_{k,\;lmn}(s,t)\;\tilde{\Gamma}_3(s,t)\;.
\end{equation}
Here the factor $\tilde{\Gamma}_3(s,t)$ is short-hand for  $\Gamma^2[-\frac{s}{2}+2k]\Gamma^2[-\frac{t}{2}+2k]\Gamma^2[-\frac{\tilde{u}}{2}+2k]$. We now let the differential operator $\Upsilon$ act on the monomial $U^{s/2+1} V^{t/2-2k+1} \sigma^m \tau^n$, leading to the following integral
\begin{equation}\label{aAUpsilonH}
\int_{-i\infty}^{i\infty} \frac{ds}{2}\; \frac{dt}{2}\; \sum_{l+m+n=k-2}a^lb^mc^n A^{s/2-2k}B^{\tilde{u}/2-2k}C^{t/2-2k}\; \Theta \;\widetilde{\mathcal{M}}_{k,\;lmn}(s,t)\; \tilde{\Gamma}_3(s,t)\;,
\end{equation}
where $U=A/B$, $V=C/B$, $\sigma=b/a$, $\tau=c/a$ have been substituted back into the expression. The factor $\Theta$ is a polynomial of $A$, $B$, $C$, $a$, $b$, $c$ and $s$, $t$ obtained from the action of the differential operator $\Upsilon$ on the monomial $U^{s/2+1} V^{t/2-2k+1} \sigma^m \tau^n$. To write  an explicit expression for $\Theta$, it is useful further to introduce the following combinations of  Mandelstam variables,
\begin{equation}
\begin{split}
X={}&s+4l-4k+2\;,\\
Y={}&t+4n-4k+2\;,\\
\tilde{Z}={}&\tilde{u}+4m-4k+2\;.
\end{split}
\end{equation}
In terms of these variables, $\Theta$ reads
\begin{equation}
\begin{split} \label{mess}
\Theta={}& -\frac{1}{4}\bigg[a b C^3 X \tilde{Z}+a B^3 c X Y+A^3 b c Y \tilde{Z}\\
+{}&A^2 b C (c (X+4) \tilde{Z}-(Y+2) \tilde{Z} (a-b+c))+A b C^2 (a (Y+4) \tilde{Z}-(X+2) \tilde{Z} (a-b+c))\\
+{}&a B C^2 (X (\tilde{Z}+2) (a-b-c)+b X (Y+4))+A^2 B c (b (4 + X) Y - (a + b - c) Y (2 + \tilde{Z}))\\
+{}&  a B^2 C ((a - b - c) X (2 + Y) + c X (4 + \tilde{Z}))\\
+{}& 
 A B^2 c ((-a - b + c) (2 + X) Y + a Y (4 + \tilde{Z}))\\
 +{}&A B C \big(a^2 (Y+2) (\tilde{Z}+2)+b^2 (X+2)( Y+2)+c^2(X+2)(\tilde{Z}+2)\\
 {}&+bc(2+X)(4+X)+ab(2+Y)(4+Y)+ac(2+\tilde{Z})(4+\tilde{Z})\big)\bigg]\;.
\end{split}
\end{equation}
The reader should not focus on this complicated expression because further manipulations will soon lead to a major simplification. At this stage we only want to point out that the above expression of $\Theta$ can be checked to be crossing invariant under any permutation of the triplets
\begin{equation}
(a,A,s)\;,\quad\quad(b,B,\tilde{u})\;,\quad\quad(c,C,t)\;.
\end{equation}
Crossing invariance of (\ref{aAUpsilonH}) implies the following crossing identities for $\widetilde{\mathcal{M}}_{k,\;lmn}(s,t)$,
\begin{equation}
\begin{split}\label{mtildecrossing}
\widetilde{\mathcal{M}}_{k,\;nml}(t,s)={}&\widetilde{\mathcal{M}}_{k,\;lmn}(s,t)\;,\\
\widetilde{\mathcal{M}}_{k,\;mnl}(\tilde{u},t)={}&\widetilde{\mathcal{M}}_{k,\;lmn}(s,t)\;,
\end{split}
\end{equation}
from which the crossing identities (\ref{crossingmtilde}) of $\widetilde{\mathcal{M}}_k(s,t;\sigma,\tau)$ immediately follow. 

As in \cite{Rastelli:2016nze,longads5}, we should reinterpret the monomials of $A$, $B$, $C$ in (\ref{mess}) as difference operators acting on functions of $s$, $t$ in the integrand, thus promoting the factor $\Theta$ to an operator $\widehat{\Theta}$. This operator $\widehat{\Theta}$ can be written in a compact form if the respective shift on $X$, $Y$ and $\tilde{Z}$ has first been performed, as we now show. All monomials that appear in $\Theta$ have the form $A^\alpha B^{3-\alpha-\beta}C^{\beta}$.
Multiplying an inverse Mellin integral by such a monomial, we have
\begin{eqnarray}
&& A^\alpha B^{3-\alpha-\beta}C^{\beta} \int_{\cal C} {ds} {dt} \, A^{s/2-2k}B^{\tilde{u}/2-2k}C^{t/2-2k} F(s,t) =  \\
&& \int_{\cal C'} ds dt A^{s/2-2k}B^{u/2-2k}C^{t/2-2k} F(s-2\alpha,t-2\beta)\;.\nonumber
\end{eqnarray}
(The shift of the integration contour is  important in producing rational terms by the mechanism discussed in \cite{longads5}. Here we are focusing on the Mellin amplitude  and  ignore contour issues.)
Note that in the first term we use  the shifted Mandelstam variable $\tilde{u} = u - 6 = 8k - s- t- 6$, while the unshifted 
$u$ appears in the second term.
We conclude that multiplication by the monomial $A^\alpha B^{3-\alpha-\beta}C^{\beta}$ corresponds in Mellin space  to a difference operator  that shifts $s\to s-2\alpha$ and $t\to t-2\beta$. 

Interpreting every monomials in $\Theta$ in this fashion we find a difference operator $\widehat{\Theta}$. We can make the expression of $\widehat{\Theta}$ very compact by performing the shift in two stages: first we shift on the factor of $X$, $Y$, $\tilde{Z}$ multiplying each monomial $A^\alpha B^{3-\alpha-\beta}C^{\beta}$ and bring it the left; then $A^\alpha B^{3-\alpha-\beta}C^{\beta}$ remains an operator to act on whatever is in the integrand on the right. We  arrive at the following simple expression,
\begin{equation} \label{Thete}
\widehat\Theta=-\frac{1}{4}\big((XY)\widehat{B\mathfrak{R}}+(XZ)\widehat{C\mathfrak{R}}+(YZ)\widehat{A\mathfrak{R}}\big)
\end{equation}
where we defined an ``unshifted'' $Z$ variable
\begin{equation}
Z\colonequals \tilde{Z}+6=u+4m-4k+2\; 
\end{equation}
and the crossing-invariant factor $\mathfrak{R}$ is given by\footnote{Curiously, this factor is closely related to
the analogous factor that $R$ that appears in the solution of the $4d$ ${\cal N}=4$ superconformal Ward identity, namely \begin{equation}
R(U,V;\sigma,\tau)=\tau \, 1+(1-\sigma-\tau)\, V+(-\tau-\sigma\tau+\tau^2)\, U+(\sigma^2-\sigma-\sigma\tau)\, UV+ \sigma V^2+\sigma\tau \, U^2 \, ,
\end{equation}  by $\mathfrak{R}=a^2B^2$. We don't have a deep understanding of this observation.
}
\begin{equation}
\mathfrak{R}=a^2 B C+ b^2AC+ c^2A B+a b C (-A-B+C)+ac B (-A+B-C)+b c A (A-B-C)
\end{equation}
The expressions $\widehat{A\mathfrak{R}}$, $\widehat{B\mathfrak{R}}$, $\widehat{C\mathfrak{R}}$ are written shorthands that should be understood as follows: one first expands $A\mathfrak{R}$, $B\mathfrak{R}$, $C\mathfrak{R}$ into monomials $A^\alpha B^{3-\alpha-\beta}C^{\beta}$ and then regards each of them as the operator $\widehat{A^\alpha B^{3-\alpha-\beta}C^{\beta}}$.  These operators will only act on objects multiplied from the right and will no longer shift the $X$, $Y$, $Z$ factors multiplied from the left.

 We can now give the explicit action of $\widehat{A^\alpha B^{3-\alpha-\beta}C^{\beta}}$ as an operator that transforms a term of $\widetilde{\mathcal{M}}_k$ into a term of $\mathcal{M}_k$,
\begin{equation}\label{monomial}
\begin{split}
\widehat{A^\alpha B^{3-\alpha-\beta}C^\beta} {}&\circ  \widetilde{\mathcal{M}}_{k,\;lmn}(s,t)\colon=  \widetilde{\mathcal{M}}_{k,\;lmn}(s-2\alpha,t-2\beta)\\
\times {}&\frac{\Gamma^2[-\frac{s}{2}+2k+\alpha]\Gamma^2[-\frac{t}{2}+2k+\beta]\Gamma^2[-\frac{u}{2}+2k+(3-\alpha-\beta)]}{\Gamma^2[-\frac{s}{2}+2k]\Gamma^2[-\frac{t}{2}+2k]\Gamma^2[-\frac{u}{2}+2k]}\;.
\end{split}
\end{equation}
This action is  obtained by applying the aforementioned shift of $s$ and $t$ on the integrand and taking into consideration the difference of Gamma function factors between the definitions (\ref{mellinG}) and (\ref{aAUpsilonH}).

To summarize, the superconformal Ward identity implies that the full Mellin amplitude $\mathcal{M}_k$ can be written in terms of an auxiliary amplitude $\widetilde{\mathcal{M}}_k$ acted upon by the difference operator $\widehat{\Theta}$,
\begin{equation}\label{prescription}
\mathcal{M}_k=\widehat{\Theta}\circ\widetilde{\mathcal{M}}_k\;.
\end{equation}
The operator $\widehat{\Theta}$ is given by  (\ref{Thete}) where each monomial operator acts as in (\ref{monomial}).

\subsection{Consistency conditions and the algebraic problem}\label{algeprob}

We  now take stock and summarize the conditions on the Mellin amplitude that follow from our discussion in the previous sections:
\begin{enumerate}
\item {\it Crossing symmetry:}  As the external operators are identical bosonic operators, the Mellin amplitude ${\mathcal{M}}_k$ satisfies the crossing relations
\begin{equation}
\begin{split}
\sigma^{k} {\mathcal{M}}_k( u,t;1/\sigma,\tau/\sigma)={}& {\mathcal{M}}_k(s,t,;\sigma,\tau)\\
\tau^{k} {\mathcal{M}}_k(t,s;\sigma/\tau,1/\tau)={}& {\mathcal{M}}_k(s,t;\sigma,\tau) \,.
\end{split}
\end{equation}

\item {\it Analytic properties:}  $\mathcal{M}_k$ has only simple poles in correspondence with the exchanged single-trace operators. Denoting the position of the simple poles in the s-, t- and u-channel as $s_0$, $t_0$, $u_0$, they are:
\begin{equation}
\begin{split}
s_0={}&4,6,\ldots, 4k-2\;,\\
t_0={}&4,6,\ldots, 4k-2\;,\\
u_0={}&4,6,\ldots, 4k-2\;.
\end{split}
\end{equation}
Moreover, the residue at any of the poles must be a polynomial in the other Mandelstam variable.

\item {\it Asymptotic behavior:}  $\mathcal{M}_k$ should grow at most linearly in the asymptotic regime of large Mandelstam variables,
\begin{equation}
\mathcal{M}_k(\beta s,\beta t,\beta u,\sigma,\tau)\sim O(\beta)\;,\quad\quad\quad\beta\to\infty\;.
\end{equation}

\item{\it Superconformal symmetry:}  $\mathcal{M}_k$ can be written in terms an auxiliary amplitude  $\widetilde{\mathcal{M}}_k$ acted upon by the difference operator $\widehat \Theta$,
\begin{equation}\label{prescription}
\mathcal{M}_k=\widehat{\Theta}\circ\widetilde{\mathcal{M}}_k\; ,
\end{equation}
where the action of $\widehat \Theta$ has been defined in the previous subsection.

\end{enumerate}
The name of the game is  to find a function $\widetilde{\mathcal{M}}_k (s, t ; \sigma, \tau)$ such that all the conditions are simultaneously satisfied. We leave a detailed analysis of this very constrainted ``bootstrap'' problem for the future.
As in the four-dimensional case analyzed in  \cite{Rastelli:2016nze,longads5}, we find it very plausible that 
this problem  has a unique solution (up to overall rescaling).

\subsection{Solutions from the position space method}\label{mamplitudes}
In this subsection we give  solutions to the algebraic problem defined above for $k=2, 3, 4$. These solutions are obtained from the position space results, but look much simpler in Mellin space when  the  prescription (\ref{prescription}) is implemented.

\subsubsection*{$k=2$}
We start from the simplest example of $k=2$. In this case, the homogenous part $\mathcal{H}_2$ is a degree-0 polynomial of $\sigma$ and $\tau$. Therefore there is only one R-symmetry structure in the auxiliary amplitude $\widetilde{\mathcal{M}}_2$. The answer is given by
\begin{equation}\label{mtildekeq2}
	\widetilde{\mathcal{M}}_2(s,t;\sigma,\tau)= \frac{32}{n^3(s-6)(s-4)(t-6)(t-4)(\tilde{u}-6)(\tilde{u}-4)}\;
	\end{equation}
which is manifestly symmetric under the permutation of $s$, $t$ and $\tilde{u}$. 	
	
\subsubsection*{$k=3$}
Moving on to the next simplest case of $k=3$, we know that $\mathcal{H}_3$ is a degree-one polynomial of $\sigma$ and $\tau$ and therefore consists of three terms. The three R-symmetry monomials $\sigma$, $\tau$, $1$ are in the same orbit under the action of the crossing symmetry group. Hence, $\widetilde{\mathcal{M}}_{3,010}$ and $\widetilde{\mathcal{M}}_{3,001}$ are related to $\widetilde{\mathcal{M}}_{3,100}$ via (\ref{mtildecrossing})
\begin{equation}
\widetilde{\mathcal{M}}_{3,010}(s,t)=\widetilde{\mathcal{M}}_{3,100}(\tilde{u},t),\;\;\;\;\;\;\;\;\;\; \widetilde{\mathcal{M}}_{3,001}(s,t)=\widetilde{\mathcal{M}}_{3,100}(t,s).
\end{equation}
Finally, $\widetilde{\mathcal{M}}_{3,100}$ is given by
\begin{equation}
\widetilde{\mathcal{M}}_{3,100}(s,t)=\frac{8(s-7)}{3n^3(s-8)(s-6)(s-4)(t-10)(t-8)(\tilde{u}-10)(\tilde{u}-8)}\;,
\end{equation}
and the full auxiliary Mellin amplitude is 
\begin{equation}
a \widetilde{\mathcal{M}}_3(s,t;\sigma,\tau)=a\widetilde{\mathcal{M}}_{3,100}(s,t)+b\widetilde{\mathcal{M}}_{3,010}(s,t)+c\widetilde{\mathcal{M}}_{3,001}(s,t)\;.
\end{equation}

\subsubsection*{$k=4$}
$\mathcal{H}_4$ is a degree-2 polynomial of $\sigma$ and $\tau$ and therefore the auxiliary amplitude contains six R-symmetry monomials. Crossing symmetry groups these six terms into two orbits $(1,\sigma^2,\tau^2)$ and $(\sigma,\tau,\sigma\tau)$, within in each amplitudes are related by (\ref{mtildecrossing}),
\begin{equation}
\begin{split}
\widetilde{\mathcal{M}}_{4,200}(s,t)={}&\widetilde{\mathcal{M}}_{4,020}(\tilde{u},t)=\widetilde{\mathcal{M}}_{4,002}(t,s)\;,\\
\widetilde{\mathcal{M}}_{4,011}(s,t)={}&\widetilde{\mathcal{M}}_{4,101}(\tilde{u},t)=\widetilde{\mathcal{M}}_{4,110}(t,s)\;.
\end{split}
\end{equation}
The two independent auxiliary amplitudes turn out to be
\begin{equation}
\begin{split}
\widetilde{\mathcal{M}}_{4,200}(s,t)={}&\frac{1}{7425n^3}\prod_{i=2}^7\frac{1}{s-2i}\prod_{j=6}^7\frac{1}{t-2j}\prod_{k=6}^7\frac{1}{\tilde{u}-2k}\\
\times {}&  \left(165 s^4-6820 s^3+102620 s^2-661648 s+1525632\right)\;,\\
\widetilde{\mathcal{M}}_{4,101}(s,t)={}&\frac{4}{7425n^3} \prod_{i=4}^7\frac{1}{s-2i}\prod_{j=4}^7\frac{1}{t-2j}\prod_{k=6}^7\frac{1}{\tilde{u}-2k}\\
\times {}& (165 s^2 t^2-4180 s^2 t+26180 s^2-4180 s t^2+105980 s t\\
{}&-664424 s+26180 t^2-664424 t+4170432)\;
\end{split}
\end{equation}
and the full auxiliary amplitude is
\begin{equation}
\begin{split}
a^2 \widetilde{\mathcal{M}}_4(s,t;\sigma,\tau)={}&a^2\widetilde{\mathcal{M}}_{4,200}(s,t)+b^2\widetilde{\mathcal{M}}_{4,020}(s,t)+c^2\widetilde{\mathcal{M}}_{4,002}(s,t)\\
+{}& ab\widetilde{\mathcal{M}}_{4,110}(s,t)+bc\widetilde{\mathcal{M}}_{4,011}(s,t)+ac\widetilde{\mathcal{M}}_{4,101}(s,t)
\end{split}
\end{equation}
The above expression  is much more complicated than those for $k=2$ and $k=3$, but the auxiliary amplitude is still significantly simpler than the full amplitude. We have not yet been able to discern a pattern that would allow to guess the auxiliary amplitude
for general $k$.

\section*{Aknowledgement}
We thank Eric Perlmutter for helpful discussions on the $\mathcal{W}_n$ four-point functions. Our work is supported in part by NSF Grant PHY-1620628.

\appendix
\section{R-symmetry polynomials}\label{Rpoly}
We collect here the R-symmetry polynomials  \cite{Nirschl:2004pa} needed for the position space method:
\begin{equation}
\begin{split}
Y_{00}=&1\;,\\
Y_{10}=& \sigma-\tau\;,\\
Y_{11}=&\sigma-\tau-\frac{2}{d}\;,\\
Y_{20}=&\sigma^2+\tau^2-2\sigma\tau-\frac{2}{d-2}(\sigma+\tau)+\frac{2}{(d-2)(d-1)}\;,\\
Y_{21}=&\sigma^2-\tau^2-\frac{4}{d+2}(\sigma-\tau)\;,\\
Y_{22}=&\sigma^2+\tau^2+4\sigma\tau-\frac{8}{d+4}(\sigma+\tau)+\frac{8}{(d+2)(d+4)}\;,\\
Y_{30}=&\sigma^3-3\sigma^2\tau+3\sigma\tau^2-\tau^3-\frac{6}{d}(\sigma^2-\tau^2)+\frac{12}{d(d+1)}(\sigma-\tau)\;,\\
Y_{31}=&\sigma^3-\sigma^2\tau-\sigma\tau^2+\tau^3-\frac{8(d-1)}{(d+4)(d-2)}(\sigma^2+\tau^2)+\frac{8(d-6)}{(d+4)(d-2)}\sigma\tau\\
{}&+\frac{4(3d+2)}{(d+1)(d+4)(d-2)}(\sigma+\tau)-\frac{8}{(d+1)(d+4)(d-2)}\;,\\
Y_{32}=&\sigma^3+3\sigma^2\tau-3\sigma\tau^2-\tau^3-\frac{12}{d+6}(\sigma^2-\tau^2)+\frac{24}{(d+4)(d+6)}(\sigma-\tau)\;,\\
Y_{33}=&\sigma^3+9\sigma^2\tau+9\sigma\tau^2+\tau^3-\frac{18}{d+8}(\sigma^2+\tau^2)-\frac{72}{d+8}\sigma\tau\\
{}&+\frac{72}{(d+6)(d+8)}(\sigma+\tau)-\frac{48}{(d+4)(d+6)(d+8)}\;.\\
\end{split}
\end{equation}
In the above expressions $d \equiv 5$. The $USp(4)$ 
 Dynkin labels $[2(a-b),2b]$ are related to the labels $(m,n)$ in $Y_{nm}$ via 
$ n=a$, $m=b$.

\section{Fixing the overall constant}\label{lastcoe}
In Section \ref{pspace} we have shown in examples how the superconformal Ward identity determines all the coefficients of the position space ansatz,  up to an overall scaling  factor $\xi$. To fix the overall normalization, we 
are going to demand that the  singular terms of the four-point function that correspond to exchanged one-half BPS operators in (say) the s-channel OPE have the correct coefficients. Those coefficients are read off from the one-half BPS
three-point functions, 
which have been computed both in supergravity  \cite{Corrado:1999pi, Bastianelli:1999en} 
and using the $W_\infty$ algebra \cite{Beem:2014kka},
\begin{equation}\label{3ptcoeff}
\begin{split}
{}&\langle {\cal O}^{(k_1)}_{\alpha_1}(x_1)   {\cal O}^{(k_2)}_{\alpha_2}(x_2)  {\cal O}^{(k_3)}_{\alpha_3}(x_3)    \rangle =\frac{C(k_1,k_2,k_3)}{x_{12}^{2k_{123}}x_{23}^{2k_{231}}x_{31}^{2k_{312}}}\langle C^{\alpha_1}C^{\alpha_2}C^{\alpha_3}\rangle\;,\\
{}&C(k_1,k_2,k_3)=\frac{2^{2\beta-2}}{(\pi n)^{\frac{3}{2}}}\Gamma\bigg(\frac{k_1+k_2+k_3}{2}\bigg)\frac{\Gamma(\frac{k_{123}+1}{2})\Gamma(\frac{k_{231}+1}{2})\Gamma(\frac{k_{312}+1}{2})}{\sqrt{\Gamma(2k_1-1)\Gamma(2k_2-1)\Gamma(2k_3-1)}}\;,\\
{}& k_{ijk}\colon= k_i+k_j-k_k\;,\\
{}& \beta= \frac{k_1+k_2+k_3}{2}\;,
\end{split}
\end{equation}
where unit normalized two-point functions are assumed. Here $C_{\alpha_i}^{I_{1} \ldots I_{k_i}}$ form a basis of symmetric traceless tensors of the R-symmetry group and we labeled such representations with $\alpha_i$. The factor $\langle C_{\alpha_1}C_{\alpha_2}C_{\alpha_3}\rangle$ is the unique scalar contraction of $C_{\alpha_i}^{I_{1}\dots I_{k_i}}$. More details on the $C$-algebra can be found in, {\it e.g.}, \cite{Lee:1998bxa}.

To keep track of various normalization factors, we start recalling that a scalar field $\phi$ with canonically normalized action
\begin{equation}
S=\int\sqrt{g}\frac{1}{2}(\partial^\mu\phi\partial_\mu\phi+m^2\phi^2)
\end{equation}
has the following bulk-to-boundary propagator
\begin{equation}
G_{B\partial}^\Delta(z,\vec{x})=\frac{\Gamma(\Delta)}{\pi^{d/2}\Gamma(\Delta-d/2)}\bigg(\frac{z_0}{z_0^2+(\vec{z}-\vec{x})^2}\bigg)^\Delta\; ,
\end{equation}
where $\Delta (\Delta - d) = m^2$.
In this normalization, the two-point function of the dual operator $\tilde {\mathcal{O}}_\Delta$ is \cite{Freedman:1998tz}) 
\begin{equation}
\langle  \tilde {  \mathcal{O}}_\Delta(\vec{x})  \tilde {\mathcal{O}}_\Delta(\vec{y})\rangle=\frac{\Gamma(\Delta)(2\Delta-d)}{\pi^{d/2}\Gamma(\Delta-d/2)}\frac{1}{(\vec{x}-\vec{y})^{2\Delta}} \, .
\end{equation}
Assuming a cubic bulk vertex of the form
\begin{equation}
\int  \frac{\mu_{\Delta\Delta\Delta'}}{2}\phi\phi\phi'   
\end{equation}
the three-point function is
\begin{equation}
\langle{\tilde{\mathcal{O}}}_\Delta(\vec{x_1}){\tilde{\mathcal{O}}}_\Delta(\vec{x_2})\tilde{\mathcal{O}}_{\Delta'}(\vec{x_3})\rangle=\frac{\mu_{\Delta\Delta\Delta'}}{x_{12}^{\Delta_{123}}x_{23}^{\Delta_{231}}x_{31}^{\Delta_{312}}}
\bigg(-\frac{\Gamma(\Delta-\Delta'/2)\Gamma^2(\Delta'/2)\Gamma((2\Delta+\Delta'-d)/2)}{2\pi^d\Gamma^2(\Delta-d/2)\Gamma(\Delta'-d/2)}\bigg)\,
\end{equation}
Finally the four-point function for the  s-channel exchange of  $\mathcal{O}_{\Delta'}$   
\begin{equation}
\langle\tilde{\mathcal{O}}_\Delta(\vec{x}_1)\tilde{\mathcal{O}}_\Delta(\vec{x}_2)\tilde{\mathcal{O}}_\Delta(\vec{x}_3)\tilde{\mathcal{O}}_\Delta(\vec{x}_4)\rangle_{\Delta'}=\mu_{\Delta\Delta\Delta'}^2\bigg(\frac{\Gamma(\Delta)}{\pi^{d/2}\Gamma(\Delta-d/2)}\bigg)^4\mathcal{E}_{\Delta'} \,.
\end{equation}
Here $\mathcal{E}_{\Delta'}$ denotes the exchange Witten diagram  for four-external scalar operators of dimension $\Delta$ and one internal scalar of dimension $\Delta'$, normalized as 
 in (A.1) of \cite{longads5}.

We now change the operator normalization such that the new operator ${\mathcal{O}}_\Delta$ is unit-normalized,
\begin{equation}
\mathcal{O}_\Delta=\bigg(\frac{\Gamma(\Delta)(2\Delta-d)}{\pi^{d/2}\Gamma(\Delta-d/2)}\bigg)^{-1/2}\tilde{{\mathcal{O}}}_\Delta \, .
\end{equation}
We specialize these formulae to our supergravity calculation. 
 We restore  the $\alpha_i$ indices and recall that the operator ${\cal O}_\alpha^{(k)}$ has dimension $\Delta = 2k$. Then
\begin{eqnarray}\label{3ptcoeffB}
\nonumber \langle{{\mathcal{O}}}^{(k)}_{\alpha_1}(\vec{x_1}){{\mathcal{O}}}^{(k)} _{\alpha_2}(\vec{x_2}){\mathcal{O}}^{(k')}_{\alpha_3}(\vec{x_3})\rangle&=&\frac{\mu_{2k,2k,2k'}}{x_{12}^{2k_{123}}x_{23}^{2k_{231}}x_{31}^{2k_{312}}} \times \bigg(\frac{\Gamma(2k)(4k-d)}{\pi^{d/2}\Gamma(2k-d/2)}\bigg)^{-3/2}\langle C_{\alpha_1}C_{\alpha_2}C_{\alpha_3}\rangle\\
&&\times \bigg(-\frac{\Gamma(2k-k')\Gamma^2(k')\Gamma((4k+2k'-d)/2)}{2\pi^d\Gamma^2(2k-d/2)\Gamma(2k'-d/2)}\bigg)
\end{eqnarray}
and 
\begin{eqnarray}\label{fourpointA}
\nonumber\langle{\mathcal{O}}^{(k)}_{\alpha_1}(\vec{x}_1){\mathcal{O}}^{(k)}_{\alpha_2}(\vec{x}_2){\mathcal{O}}^{(k)}_{\alpha_3}(\vec{x}_3){\mathcal{O}}^{(k)}_{\alpha_4}(\vec{x}_4)\rangle_{k'}&=&\mu_{2k,2k,2k'}^2\bigg(\frac{\Gamma(2k)}{\pi^{d/2}\Gamma(2k-d/2)}\bigg)^4\bigg(\frac{\Gamma(2k)(4k-d)}{\pi^{d/2}\Gamma(2k-d/2)}\bigg)^{-2}\\
&&\times \mathcal{E}_{s_{k'}} \langle C_{\alpha_1}C_{\alpha_2}C_{\alpha_5}\rangle \langle C_{\alpha_5}C_{\alpha_3}C_{\alpha_4}\rangle 
\end{eqnarray}
where the collective index $\alpha_5$ is summed over. Comparing (\ref{3ptcoeff}) with (\ref{3ptcoeffB}) solves the cubic coupling $\mu_{2k,2k,2k'}$. Inserting $\mu_{2k,2k,2k'}$ gives us the precise contribution of the exchange Witten diagram in the four-point function.

We now recall the unfixed parameter $\lambda_{s_{k'}}$ in Section \ref{pspace} (for $k'=2$, $\lambda_{s_2}\equiv\xi$) appears in the $t_i$-contracted four-point function as,
\begin{equation}
\langle{\mathcal{O}}^{(k)}(\vec{x_1},t_1){\mathcal{O}}^{(k)}(\vec{x_2},t_2){\mathcal{O}}^{(k)}(\vec{x_3},t_3){\mathcal{O}}^{(k)}(\vec{x_4},t_4)\rangle_{k'}=\lambda_{s_{k'}}\, \mathcal{E}_{s_{k'}} Y(t_1,t_2,t_3,t_4)\;.
\end{equation}
 Here the factor $Y(t_1,t_2,t_3,t_4)$ is the R-symmetry polynomial of $t_i\cdot t_j$, related to $Y_{mn}(\sigma,\tau)$ by $Y((t_1,t_2,t_3,t_4)=(t_1\cdot t_2)^k(t_3\cdot t_4)^kY_{mn}(\sigma,\tau)$. To make comparison with (\ref{fourpointA}), we only need to contract (\ref{fourpointA}) with the null vectors. Note the tensor $C_\alpha^{I_1 \ldots I_k}$ is   normalized to be
\begin{equation}
C_{\alpha_1}^{I_1 \ldots I_k}C_{\alpha_2}^{I_1 \ldots I_k}=\delta_{\alpha_1\alpha_2}
\end{equation}
with the completeness relation\footnote{Here we did not write down the extra term with mixed $i$ and $j$ because we will only need to match the $\sigma^n+\tau^n$ term.}
\begin{equation}
C_\alpha^{I_1 \ldots I_k}C_\alpha^{J_1 \ldots J_k}=\delta^{I_1\ldots I_k,(J_1\ldots J_k)}+(\text{mix within I, J}) \label{complete}\;.
\end{equation}
The null vector $t_1$ of operator 1 is contracted to $C_{\alpha_1}$ via another $C_{\alpha_1}$
\begin{equation}\label{contract1}
(t_{1,L_1}\ldots t_{1,L_p}C_{\alpha_1}^{L_1 \ldots L_p})C_{\alpha_1}^{I_1 \ldots I_p}=t_{1,I_1}\ldots t_{1,I_p}
\end{equation}
where the above completeness relation has been used. Because $t_1$ is a null vector, the implicit mixed part in the completeness relation does not contribute and  the right side of (\ref{contract1}) is exact. We contract with the null vectors $t_{2,3,4}$ in the same way.

 After some tedious algebra, we find
\begin{equation}
\xi=n_R\times \frac{32 (2k -1)^4 \Gamma (2k' ) \Gamma (2k -1)^2 \Gamma \left(\frac{1}{2} (k' +2k )\right)^2}{ n^3 \pi ^3 (2k' -2) \Gamma \left(\frac{k' }{2}\right)^4 \Gamma \left(k-\frac{k' }{2}\right)^2 \Gamma \left(k' +2k -3\right)^2}
\end{equation}
where $n_R$ is the ratio of $t_i$-contracted $\langle C_{\alpha_1}C_{\alpha_2}C_{\alpha_5}\rangle\langle C_{\alpha_5}C_{\alpha_3}C_{\alpha_4}\rangle$ and $Y(t_1,t_2,t_3,t_4)$. Note our convention for $Y(t_1,t_2,t_3,t_4)$ is always of the form 
\begin{equation}
 Y(t_1,t_2,t_3,t_4)=(t_1\cdot t_2 \;t_3\cdot t_4)^n(\sigma^n+\tau^n+\ldots) 
\end{equation}
where ``$\ldots$'' stands for the terms of the form $\sigma^a\tau^b$ with $a<n$ and $b<n$. Using the above completeness relation (\ref{contract1}),  it is not difficult to solve the combinatoric problem of $n_R$ where $C_{\alpha_{1,2,3,4}}$ have $k$ $I$-indices and $C_{\alpha_5}$ has $k'$ $I$-indices. All in all, the final answer for $n_R$ depends only on $k'$ (note due to R-symmetry selection rules, $k'$ is always an even integer)
\begin{equation}
n_R(k')=\frac{((k'/2)!)^2}{(k')!}\;.
\end{equation}

As a consistency check, let us use the above formula and compute the ratios of unfixed parameters $\lambda_{s_i}$ in the position space ansatz. For $k=3$, there are two such parameters $\lambda_{s_2}$ and $\lambda_{s_4}$. We find the ratio is 
\begin{equation}
\frac{\lambda_{s_4}}{\lambda_{s_2}}=\frac{224}{135}\;,
\end{equation}
For $k=4$ there are $\lambda_{s_2}$, $\lambda_{s_4}$ and $\lambda_{s_6}$ and we get
\begin{equation}
\frac{\lambda_{s_6}}{\lambda_{s_2}}=\frac{33}{28}\;,\quad\quad \frac{\lambda_{s_4}}{\lambda_{s_2}}=\frac{125}{42}\;.
\end{equation}
Happily, they all agree with ratios computed independently using our position space method.

\section{Four-point functions of the $\mathcal{W}_{n \to \infty}$ algebra}\label{4ptchiral}
In this Appendix we  use the  ``holomorphic bootstrap'' method of \cite{Headrick:2015gba} to compute  four-point functions of  the $\mathcal{W}_{n \to \infty}$ algebra, for the cases relevant to our supergravity computation.

We start by recalling that in a chiral  algebra the four-point function of identical quasi-primary operators, 
\begin{equation}
\langle\mathcal{O}_h(z_1)\mathcal{O}_h(z_2)\mathcal{O}_h(z_3)\mathcal{O}_h(z_4) \rangle=(z_{12}z_{34})^{-2h}\mathcal{F}(\chi)\, ,
\end{equation}
satisfies the following crossing equation
\begin{equation}
\mathcal{F}(\chi)=\chi^{2h}\mathcal{F}(1/\chi)=\chi^{2h}(1-\chi)^{-2h}\mathcal{F}(1-\chi)\;.
\end{equation}
The function $\mathcal{F}(\chi)$ can be written as a sum of the $SL(2,\mathbb{R})$ blocks
\begin{equation}
\mathcal{F}(\chi)=\sum_{i}C_{hhi}^2 \chi^{h_i}{}_2F_1(h_i,h_i;2h_i;\chi)\;.
\end{equation}
Here $C_{hhi}$ is the three-point coupling in
\begin{equation}
\langle\mathcal{O}_h(z_1)\mathcal{O}_h(z_2)\mathcal{O}_{h_i}(z_3)\rangle=\frac{C_{hhi}}{z_{12}^{2h-h_i}z_{13}^{h_i}z_{23}^{h_i}}\;,
\end{equation}
and all two-point functions are normalized to unity. Combining the crossing equation with  the conformal block decomposition, we find that the singularities of $\mathcal{F}$ at $\chi=1$ and $\chi\to\infty$ are
\begin{equation}
\begin{split}
\mathcal{F}(\chi)={}&(-1)^{4h}\sum_{i}C_{hhi}^2(1-\chi)^{h_i-2h}{}_2F_1(h_i,h_i;2h_i;1-\chi)\chi^{2h}\\
\approx{}& \sum_{n=1}^{2h}\beta_n(1-\chi)^{-n}+\ldots\quad\quad {\rm as\;} \chi\to 1\;,\\
\mathcal{F}(\chi)={}&(-1)^{4h}\sum_{i}C_{hhi}^2\chi^{2h-h_i}{}_2F_1(h_i,h_i;2h_i;1/\chi)\\
\approx{}& \sum_{n=1}^{2h}\beta_n\chi^{n}+\ldots\quad\quad {\rm as\;} \chi\to \infty\; .
\end{split}
\end{equation}
Here $\alpha_n$ and $\beta_n$ are computable numbers given the three-point functions coefficients $C_{hhi}$ with $h_i<2h$.

The meromorphic $\mathcal{F}(\chi)$ is  completely determined by these singularities and by its value at $\chi=0$,
\begin{equation}
\mathcal{F}(\chi)=1+\sum_{n=1}^{2h}\alpha_n \chi^n+\sum_{n=1}^{2h}\beta_n[(1-\chi)^{-n}-1]\;.
\end{equation}
Alternatively, we can cast $\mathcal{F}(\chi)$ into the following more convenient parameterization \cite{Headrick:2015gba},
\begin{equation}
\mathcal{F}(\chi)=\sum_{n=0}^{[2h/3]} c_n \frac{\chi^{2n}(1-\chi+\chi^2)^{2h-3n}}{(1-\chi)^{2h-2n}}\, ,
\end{equation}
which is manifestly crossing-symmetric and has the same singularity structure, where the finitely many constants $c_n$ are determined from the conformal block decomposition at $\chi=0$.

\subsection*{$k=2$}
We first look at the $k=2$ case which corresponds to $2d$ stress tensor. The general form of $\mathcal{F}(\chi)$ for $h=k=2$ reads
\begin{equation}
\mathcal{F}(\chi)=\frac{c_0 \left(\chi^2-\chi+1\right)^4+c_1 (1-\chi)^2 \chi^2 \left(\chi^2-\chi+1\right)}{(1-\chi)^4}\;.
\end{equation}
To fix the constants $c_n$ we only need to match with the OPE coefficients of $TT1$ and $TTT$
\begin{equation}
C_{TT1}=1\;,\quad\quad C_{TTT}=2^{3/2}c^{-1/2}\;.
\end{equation}
We find 
\begin{equation}
c_0=1\;, \quad\quad c_1=\frac{8}{c}-4\;.
\end{equation}
In the $1/c$ expansion, the four-point function with the above solution of  $c_n$ simply reads
\begin{equation}
\mathcal{F}(\chi)=1+\chi^4+\frac{\chi^4}{(1-\chi)^4}+\frac{1}{c}\left(\frac{8\chi^2(1-\chi+\chi^2)}{(1-\chi)^2}\right)+\mathcal{O}(c^{-2})\;.
\end{equation}
Notice the leading term is nothing but the disconnected piece of the full four-point function under the chiral algebra twist and is anticipated from the large-$c$ factorization. The subleading term in $1/c$ on the other hand reproduces precisely the holomorphic four-point function we obtained from the supergravity computation, upon recalling that $c \sim 4n^3$ in the large $n$ limit.

\subsection*{$k=3$}
In the case of $k=3$, the general form of $\mathcal{F}(\chi)$ admits three parameters,
\begin{equation}
\mathcal{F}(\chi)=\frac{c_0 \left(\chi^2-\chi+1\right)^6+c_1 (1-\chi)^2 \chi^2 \left(\chi^2-\chi+1\right)^3+c_2 (1-\chi)^4 \chi^4}{(1-\chi)^6}\;.
\end{equation}
This chiral four-point function is to be matched with the $W_3W_31$ coefficient as well as $W_3W_3T$, $W_3W_3W_4$  coefficients which are given by (\ref{3ptcoeff}). The end result is 
\begin{equation}
c_0=1\;,\quad\quad c_1=\frac{18}{c}-6\;,\quad\quad c_2=\frac{3 (c-9)}{c}\;.
\end{equation}
The large $c$ expansion of $\mathcal{F}(\chi)$ gives
\begin{equation}
\begin{split}
\mathcal{F}(\chi)={}&1+\chi^6+\frac{\chi^6}{(1-\chi)^6}\\
+{}&\frac{1}{c}\left(\frac{9 \chi ^2 \left(2 \chi ^6-6 \chi ^5+9 \chi ^4-8 \chi ^3+9 \chi ^2-6 \chi +2\right)}{ (\chi -1)^4}\right)+\mathcal{O}(c^{-2})\;,
\end{split}
\end{equation}
matching again the supergravity result.

\subsection*{$k=4$}
The general $\mathcal{F}(\chi)$ for $k=4$ can be written as
\begin{equation}
\mathcal{F}(\chi)=\frac{c_0 \left(\chi^2-\chi+1\right)^8+c_1 (1-\chi)^2 \chi^2 \left(\chi^2-\chi+1\right)^5+c_2 (1-\chi)^4 \chi^4 \left(\chi^2-\chi+1\right)^2}{(1-\chi)^8}\;.
\end{equation}
By matching with the OPE coefficients of $W_4W_41$, $W_4W_4T$, $W_4W_4W_4$, we found that
\begin{equation}
c_0=1\;,\quad\quad c_1=\frac{32}{c}-8\;,\quad\quad c_2=-\frac{672}{5c}+12\;.
\end{equation}
This leads to the following large $c$ expansion of $\mathcal{F}(\chi)$,
\begin{equation}
\begin{split}
\mathcal{F}(\chi)={}&1+\chi^8+\frac{\chi^8}{(1-\chi)^8}\\
+{}&\frac{1}{c}\left(\frac{32 \chi ^2 \left(\chi ^2-\chi +1\right)^2 \left(5 \chi ^6-15 \chi ^5+9 \chi ^4+7 \chi ^3+9 \chi ^2-15 \chi +5\right)}{5 (\chi -1)^6}\right)\\
+{}&\mathcal{O}(c^{-2})\;,
\end{split}
\end{equation}
again in perfect agreement with our supergravity computation.

\bibliography{6d} 
\bibliographystyle{utphys}

\end{document}